\documentclass[11pt]{article}

\usepackage[paper=a4paper,margin=1in]{geometry}
\usepackage{latexsym, graphicx}
\usepackage{amsmath, amssymb, bm, longtable, array, tabularx, subcaption}
\usepackage{appendix}
\usepackage{verbatim}
\usepackage{cite}
\usepackage[colorlinks=false]{hyperref}
\usepackage{placeins}

\usepackage{xcolor}

\newcommand{\fref}[1]{Figure~\ref{#1}}
\newcommand{\tref}[1]{Table~\ref{#1}}

\newcommand{\IA}{\mathbb{A}}

\newcommand{\IP}{\mathbb{P}}

\newcommand{\cicy}[2]{\begin{matrix} #1\end{matrix}\!\left[\begin{matrix}#2 \end{matrix}\right]}
\newcommand\citep{\cite}

\begin{document}
\thispagestyle{empty}
\renewcommand{\thefootnote}{\fnsymbol{footnote}}

\begin{titlepage}

${}$\\

\begin{center}

\textbf{
{\LARGE Machine Learning CICY Threefolds}
}

\vspace{0.3in}

\textbf{\large
Kieran Bull$^{a}$\footnote{\texttt{kieran.bull@seh.ox.ac.uk}},
Yang-Hui He$^{a,b,c,d}$\footnote{\texttt{hey@maths.ox.ac.uk}},
Vishnu Jejjala$^{e}$\footnote{\texttt{vishnu@neo.phys.wits.ac.za}},
Challenger Mishra$^{a,f}$\footnote{\texttt{challenger.mishra@gmail.com}}\\
}

\vspace{0.2in}

${}^a$ \textit{Rudolf Peierls Centre for Theoretical Physics, Clarendon Laboratory, \\
	Parks Rd, University of Oxford, OX1 3PU, UK}
\vskip0.25cm

${}^b$ \textit{School of Physics, NanKai University, Tianjin, 300071, P.R.\ China,}
\vskip0.25cm

${}^c$ \textit{Department of Mathematics, City, University of London, EC1V 0HB, UK}
\vskip0.25cm

${}^d$ \textit{Merton College, University of Oxford, OX1 4JD, UK}
\vskip0.25cm

${}^e$ \textit{Mandelstam Institute for Theoretical Physics, NITheP, CoE-MaSS, and School of Physics,\\
University of the Witwatersrand, Johannesburg, South Africa}
\vskip0.25cm

${}^f$ \textit{Christ Church, University of Oxford, OX1 1DP UK}

\end{center}

\vspace{0.1in}

\hyphenation{CICY}
\hyphenation{CICYs}

\begin{abstract}
\noindent The latest techniques from Neural Networks and Support Vector Machines
(SVM) are used to investigate geometric properties of Complete Intersection
Calabi--Yau (CICY)  threefolds, a class of manifolds that facilitate string
model building. An advanced neural network classifier and SVM are employed to
(1) learn Hodge numbers and report a remarkable improvement over previous
efforts, (2) query for favourability, and (3) predict discrete symmetries, a
highly imbalanced problem to which both Synthetic Minority Oversampling
Technique (SMOTE) and permutations of the CICY matrix are used to decrease the
class imbalance and improve performance.  In each case study, we employ a
genetic algorithm to optimise the hyperparameters of the neural network. We
demonstrate that our approach provides quick diagnostic tools capable of
shortlisting quasi-realistic string models based on compactification over smooth
CICYs and further supports the paradigm that classes of problems in algebraic
geometry can be machine learned.
\end{abstract}

\end{titlepage}

\newpage

\renewcommand{\thefootnote}{\arabic{footnote}}
\setcounter{footnote}{0}

\section{Introduction}\label{intro}
String theory supplies a framework for quantum gravity.
Finding our universe among the myriad of possible, consistent realisations of a four dimensional low-energy limit of string theory constitutes the vacuum selection problem.
Most of the vacua that populate the string landscape are \textit{false} in that they lead to physics vastly different from what we observe in Nature.
We have so far been unable to construct even one solution that reproduces all of the known features of particle physics and cosmology in detail.
The challenge of identifying suitable string vacua is a problem in big data that invites a machine learning approach. 

The use of machine learning to study the landscape of vacua is a relatively recent development.
Several avenues have already yielded promising results.
These include Neural Networks \cite{he2017deep, He:2017set, krefl2017machine,wang2018learning}, Linear  Regression \cite{carifio2017machine}, Logistic Regression \cite{carifio2017machine,wang2018learning}, Linear Discriminant Analysis, k-Nearest Neighbours, Classification and Regression Tree, Naive Bayes \cite{carifio2017machine}, Support Vector Machines \cite{carifio2017machine,wang2018learning}, Evolving Neural Networks \cite{ruehle2017evolving}, Genetic Algorithms \cite{abel2014genetic}, Decision Trees and Random Forest \cite{wang2018learning}, Network Theory \cite{Carifio:2017nyb}. 

Calabi--Yau threefolds occupy a central r\^ole in the study of the string
landscape.  In particular, Standard Model like theories can be engineered from
compactification on these geometries.  As such, Calabi--Yau manifolds have been
the subject of extensive study over the past three decades.  Vast datasets of
their properties have been constructed, warranting a deep-learning approach
\cite{he2017deep, He:2017set}, wherein a paradigm of machine learning
computational algebraic geometry has been advocated.  In this paper, we employ
feedforward neural networks and support vector machines to probe a subclass of
these manifolds to extract topological quantities.  We summarise these
techniques below.  \begin{itemize}
\item Inspired by their biological counterparts, artificial \textbf{Neural Networks} constitute a class of machine learning techniques capable of dealing with both classification and regression problems.
In practice, they can be thought of as highly complex functions acting on an input vector to produce an output vector.
There are several types of neural networks, but in this work we employ feedforward neural networks, wherein information moves in the forward direction from the input nodes to the output nodes via hidden layers. 
We provide a brief overview of feedforward neural networks in Appendix~\ref{NeuralNetworks}.
\item \textbf{Support Vector Machines (SVMs)}, in contrast to neural networks, take a more geometric approach to machine learning.
SVMs work by constructing hyperplanes that partition the feature space and can be adapted to act as both classifiers and regressors.
A brief overview is presented in Appendix~\ref{SVM}.
\end{itemize}
The manifolds of interest to us are the Complete Intersection Calabi--Yau threefolds (CICYs), which we review in the following section.
The CICYs generalise the famous quintic as well as Yau's construction of the Calabi--Yau threefold embedded in $\IP^3{\times}\IP^3$ \cite{green1987calabi}.
The simplicity of their description makes this class of geometries particularly amenable to the tools of machine learning.
The choice of CICYs is however mainly guided by other considerations.
First, the CICYs constitute a sizeable collection of Calabi--Yau manifolds and are in fact the first such large dataset in algebraic geometry.
Second, many properties of the CICYs have already been computed over the years, like their Hodge numbers \cite{green1987calabi, candelas1988complete} and discrete isometries \cite{Candelas:2008wb, free_syms, lukas2017discrete, candelas2018highly}.
The Hodge numbers of their quotients by freely acting discrete isometries have also been computed \cite{Candelas:2008wb, Candelas:2010ve, candelas2016hodge, constantin2017hodge, constantin2016calabi}.
In addition, the CICYs provide a playground for string model building.
The construction of stable holomorphic vector \cite{Anderson:2007nc, Anderson:2008uw, anderson2010exploring, Anderson:2011ns,  Anderson:2012yf} and monad bundles \cite{anderson2010exploring} over smooth favourable CICYs has produced several quasi-realistic heterotic string derived Standard Models through intermediate GUTs.
These constitute another large dataset based on these manifolds. 

Furthermore, the Hodge numbers of CICYs were recently shown to be machine learnable to a reasonable degree of accuracy using a primitive neural network of the multi-layer perceptron type \cite{he2017deep}.
In this paper, we consider whether a more powerful machine learning tool (like a more complex neural network) or an SVM yields significantly better results.
We wish to learn the extent to which such topological properties of CICYs are machine learnable, with the foresight that machine learning techniques can become a powerful tool in constructing ever more realistic string models, as well as helping understand Calabi--Yau manifolds in their own right. 

Guided by these considerations, we conduct three case studies over the class of CICYs.
We first apply SVMs and neural networks to machine learn the Hodge number $h^{1,1}$ of CICYs.
We then attempt to learn whether a CICY is favourably embedded in a product of projective spaces, and whether a given CICY admits a quotient by a freely acting discrete symmetry. 

The paper is structured as follows.
In Section~\ref{datasets}, we provide a brief overview of CICYs and the datasets over them relevant to this work. 
In Section~\ref{benchmark}, we discuss the metrics for our machine learning paradigms.
Finally, in Section~\ref{case_studies}, we present our results.

\section{The CICY Dataset}\label{datasets}
A CICY threefold is a Calabi--Yau manifold embedded in a product of complex projective spaces, referred to as the ambient space.
The embedding is given by the zero locus of a set of homogeneous polynomials over the combined set of homogeneous coordinates of the projective spaces.
The deformation class of a CICY is then captured by a \textit{configuration matrix} \eqref{eq:confmatr1}, which collects the multi-degrees of the polynomials:
\begin{equation}\label{eq:confmatr1}
X~=~~
\cicy{\IP^{\,n_1} \\[4pt] \vdots\\[4pt] \IP^{\,n_m}}
{ ~q^1_1 & &\ldots && q^1_K \\[7pt]
  ~\vdots & &\ddots && \vdots \\[4pt]
  ~q^m_1 & &\ldots && q^m_K} , \quad q^r_a\in \mathbb{Z}_{\ge 0} .
\end{equation}\\[0.5pt]
In order for the configuration matrix in \eqref{eq:confmatr1} to describe a CICY threefold, we require that ${\sum}_r n_r - K = 3$.
In addition, the vanishing of the first Chern class is accomplished by demanding that $\sum_{a} q^r_a = n_r+1$, for each $r\in\{1,\ldots,m\}$.
There are $7890$ CICY configuration matrices in the CICY list (available online at \cite{cicylist2}).
At least $2590$ of these are known to be distinct as classical manifolds. 

The Hodge numbers $h^{p,q}$ of a Calabi--Yau manifold are the dimensions of its Dolbeault cohomology classes $H^{p,q}$.
A related topological quantity is the Euler characteristic $\chi$.
We define these quantities below:
\begin{equation}
h^{p,q}=\text{dim}~{H^{p,q}} \text{,~~~~~~}\chi=\sum_{p,q=0}^3 (-1)^{p+q} h^{p,q}\text{,~~~~~~}p,q\in\{0,1,2,3\}
\end{equation}
For a smooth and connected Calabi--Yau threefold with holonomy group $SU(3)$, the only unspecified Hodge numbers are $h^{1,1}$ and $h^{2,1}$.
These are topological invariants that capture the dimensions of the K\"ahler and the complex structure moduli spaces, respectively.
The Hodge numbers of all CICYs are readily accessible \cite{cicylist2}.
There are $266$ distinct Hodge pairs $(h^{1,1},h^{2,1})$ of the CICYs, with $0\le{h^{1,1}}\le 19$ and $0\le{h^{2,1}}\le 101$.
From a model building perspective, knowledge of the Hodge numbers is imperative to the construction of a string derived Standard Model. 

If the entire second cohomology class of the CICY descends from that of the ambient space ${\IA}={\IP}^{n_1}{\times}{\ldots}{\times}\, \IP^{n_m}$, then we identify the CICY as \textit{favourable}.
There are $4874$ favourable CICYs \cite{cicylist2}.
As an aside, we note that it was shown recently that all but $48$ CICY configuration matrices can be brought to a favourable form through \textit{ineffective splittings} \cite{anderson2017fibrations}.
The remaining can be seen to be favourably embedded in a product of del Pezzo surfaces.
The favourable CICY list is also available online \cite{favcicylist}.
(In this paper, we will not be concerned with this new list of CICY configuration matrices.)
The favourable CICYs have been especially amenable to the construction of stable holomorphic vector and monad bundles, leading to several quasi-realistic heterotic string models.

Discrete symmetries are one of the key components of string model building.  The
breaking of the GUT group to the Standard Model gauge group proceeds via
discrete Wilson lines, and as such requires a non-simply connected
compactification space.  Prior to the classification efforts
\cite{Candelas:2008wb, free_syms}, almost all known Calabi--Yau manifolds were
simply connected.  The classification resulted in identifying all CICYs that
admit a quotient by a freely acting symmetry, totalling $195$ in number, $2.5$\%
of the total, creating a highly unbalanced dataset.  $31$ distinct symmetry
groups were found, the largest being of order $32$.  Counting inequivalent
projective representations of the various groups acting on the CICYs, a total of
$1695$ CICY quotients were obtained \cite{cicylist2}. 

A CICY quotient might admit further discrete symmetries that survive the breaking of the string gauge group to the Standard Model gauge group.
These in particular are phenomenologically interesting since they may address questions related to the stability of the proton via $R$-symmetries and the structure of the mixing matrices via non-Abelian discrete symmetries.
A classification of the remnant symmetries of the $1695$ CICY quotients found that $381$ of them had nontrivial remnant symmetry groups \cite{lukas2017discrete}, leading to a more balanced dataset based on symmetry.
We will however focus on the first symmetry dataset available at \cite{cicylist2} purely on the grounds that the size of the dataset is itself much larger than the latter dataset. 

\section{Benchmarking Models} \label{benchmark}
In order to benchmark and compare the performance of each machine learning approach we adapt in this work, we use {cross validation} and a range of other statistical measures.
Cross validation means we take our entire data set and split it into training and validation sets.
The training set is used to train models whereas the validation set remains entirely unseen by the machine learning algorithm.
Accuracy measures computed on the training set thus give an indication of the model's performance in recalling what it has learned.
More importantly, accuracy measures computed against the validation set give an indication of the models performance as a predictor. 

For regression problems we make use of both root mean square error (RMS) and the coefficient of determination ($R^2$)
to assess performance:
\begin{align}
	\text{RMS} & := \left({ \frac{1}{N} \sum_{i=1}^N (y^{pred}_i-y_i)^2 })\right)^{1/2}, \quad
	R^2  := 1 - \frac{\sum_i (y_i-y^{pred}_i)^2}{\sum_i (y_i-\bar{y})^2}\,,
\end{align}
where $y_i$ and $y_i^{pred}$ stand for actual and predicted values, with $i$ taking values in 1 to $N$, and $\bar{y}$ stands for the average of all $y_i$.
A rudimentary binary accuracy is also computed by rounding the predicted value and counting the results in agreement with the data.
As this accuracy is a binary success or failure, we can use this measure to calculate a Wilson confidence interval. Define
\begin{align}
\omega_\pm:=	\frac{p+\frac{z^2}{2n}}{1+\frac{z^2}{n}}
	\pm \frac{z}{1+ \frac{z^2}{n}} \left({
	\frac{p(1-p)}{n} + \frac{z^2}{4 n^2} }\right)^{1/2}\,,
\end{align}
where $p$ is the probability of a successful prediction, $n$ the number of entries in the dataset, and $z$ the \textit{probit} of the normal distribution (\textit{e.g.}, for a $99\%$ confidence interval, $z=2.575829$).
The upper and the lower bounds of this interval are denoted by WUB and WLB respectively.  

For classifiers, we have addressed two distinct types of problems in this paper, namely balanced and imbalanced problems.
Balanced problems are where the number of elements in the true and false classes are comparable in size.
Imbalanced problems, or the so called \textit{needle in a haystack}, are the opposite case.
It is important to make this distinction, since models trained on imbalanced problems can easily achieve a high accuracy, but accuracy would be a meaningless metric in this context.
For example, consider the case where only $\sim 0.1\%$ of the data is classified as true.
In minimising its cost function on training, a neural network could naively train a model which just predicts false for any input.
Such a model would achieve a $99.9\%$ accuracy, but it is useless in finding the special few cases that we are interested in.
A different measure is needed in these cases.
For classifiers, the possible outcomes are summarised by the \textit{confusion matrix} of \tref{confusionmatrix}, whose elements we use to define several performance metrics:
\begin{align}
	\text{TPR} := \frac{tp}{tp+fn}&,& ~~~~&\text{FPR} := \frac{fp}{fp+tn} \,,\\
	\text{Accuracy} := \frac{tp+tn}{tp+tn+fp+fn}\,&,&~~~ &\text{Precision} := \frac{tp}{tp+fp}\,. \nonumber
\end{align}
where, TPR (FPR) stand for True (False) Positive Rate, the former also known as \textit{recall}.
\begin{table}
\begin{center}
\small
\begin{tabular}{cc|c|c|}
	\cline{3-4} 
	&  & \multicolumn{2}{c|}{{Actual}} \\ \cline{3-4}
	& &  True & False \\ \hline
	\multicolumn{1}{|c|}{{Predicted}} & True & True Positive
	($tp$) & False Positive ($fp$) \\  \cline{2-4}
	\multicolumn{1}{|c|}{{Classification}} & False & False Negative
	($fn$)& True Negative ($tn$)\\ \hline
\end{tabular}
	\caption{Confusion matrix.}\label{confusionmatrix}
\end{center}
\normalsize
\end{table}
For balanced problems, accuracy is the go-to performance metric, along with its associated Wilson confidence interval. However, for imbalanced problems, we use $F$-values and AUC.
We define,
\begin{align}
	F:= \frac{2}{ \frac{1}{\text{Recall}} + \frac{1}{\text{Precision}}  }\,,
\end{align}
while AUC is the area under the \textit{receiver operating characteristic} (ROC) curve that plots TPR against FPR.
$F$-values vary from $0$ to $1$, whereas AUC ranges from $0.5$ to $1$.
We will discuss these in greater detail in Section \ref{s:ROC}.

\section{Case Studies}\label{case_studies}
We conduct three case studies over the CICY threefolds.
Given a CICY threefold $X$, we explicitly try to learn the topological quantity $h^{1,1}(X)$, the Hodge number that captures the dimension of the K\"ahler structure moduli space of $X$.
We then attempt a (balanced) binary query, asking whether a given manifold is favourable.
Finally, we attempt an (imbalanced) binary query about whether a CICY threefold X, admits a quotient $X/G$ by a freely acting discrete isometry group $G$.
\subsection{Machine Learning Hodge Numbers} \label{hodge}
As noted in Section~\ref{datasets}, the only independent Hodge numbers of a Calabi--Yau threefold are $h^{1,1}$ and $h^{2,1}$.
We attempt to machine learn these.
For a given configuration matrix \eqref{eq:confmatr1} describing a CICY, the Euler characteristic $\chi = 2(h^{1,1}-h^{2,1})$ can be computed from a simple combinatorial formula \citep{bestiary}.
Thus, it is sufficient to learn only one of the Hodge numbers.
We choose to learn $h^{1,1}$ since it takes values in a smaller range of integers than $h^{2,1}$.
\subsubsection{Architectures}
To determine the Hodge numbers we use regression machine learning techniques to predict a continuous output with the CICY configuration matrix \eqref{eq:confmatr1} as the input.
The optimal SVM hyperparameters were found by hand to be a Gaussian kernel with $\sigma{=}2.74$, $C{=}10$, and $\epsilon{=}0.01$.
Optimal neural network hyperparameters were found with a genetic algorithm, leading to an overall architecture of five hidden layers with $876$, $461$, $437$, $929$, and $404$ neurons, respectively.
The algorithm also found that a ReLU (rectified linear unit) activation layer and dropout layer of dropout $0.2072$ between each neuron layer give optimal results.
(See Appendix \ref{hyper} for a description of hyperparameters.)

A neural network classifier was also used.
To achieve this, rather than using one output layer as is the case for a binary
classifier or regressor, we use an output layer with $20$ neurons (since $h^{1,1} \in (0,19)$) with each neuron mapping to $0/1$, the location of the $1$ corresponding to a unique $h^{1,1}$ value.
Note this is effectively adding extra information to the input as we are explicitly fixing the range of allowed $h^{1,1}$s.
For a large enough training data size this is not an issue, as we could extract this information from the training data (choose the output to be the largest $h^{1,1}$ from the training data --- for a large enough sample it is likely to contain $h^{1,1} = 19$).
Moreover, for a small training data size, if only $h^{1,1}$ values less than a given number are present in the data, the model will not be able to learn these $h^{1,1}$ values anyway --- this would happen with a continuous output regression
model as well.

The genetic algorithm is used to find the optimal classifier architecture.
Surprisingly, it finds that adding several {convolution layers} led to the best
performance.  This is unexpected as convolution layers look for features which
are translationally or rotationally invariant (for example, in number
recognition they may learn to detect rounded edges and associate this with a
zero).  Our CICY configurations matrices do not exhibit these symmetries, and
this is the only result in the paper where convolution layers lead to better
results rather than worse.  The optimal architecture was found to be be four
convolution layers with $57$, $56$, $55$, and $43$ feature maps, respectively,
all with a kernel size of $3{\times}3$.  These layers were followed by two
hidden fully connected layers and the output layer, the hidden layers containing
$169$ and $491$ neurons.  ReLU activations and a dropout of $0.5$ were included
between every layer, with the last layer using a sigmoid activation. Training
with a laptop computer's CPU took less than 10 minutes and execution on the
validation set after training takes seconds.

\begin{figure}[h!!]
	\centering
		\includegraphics[width=0.67\textwidth]{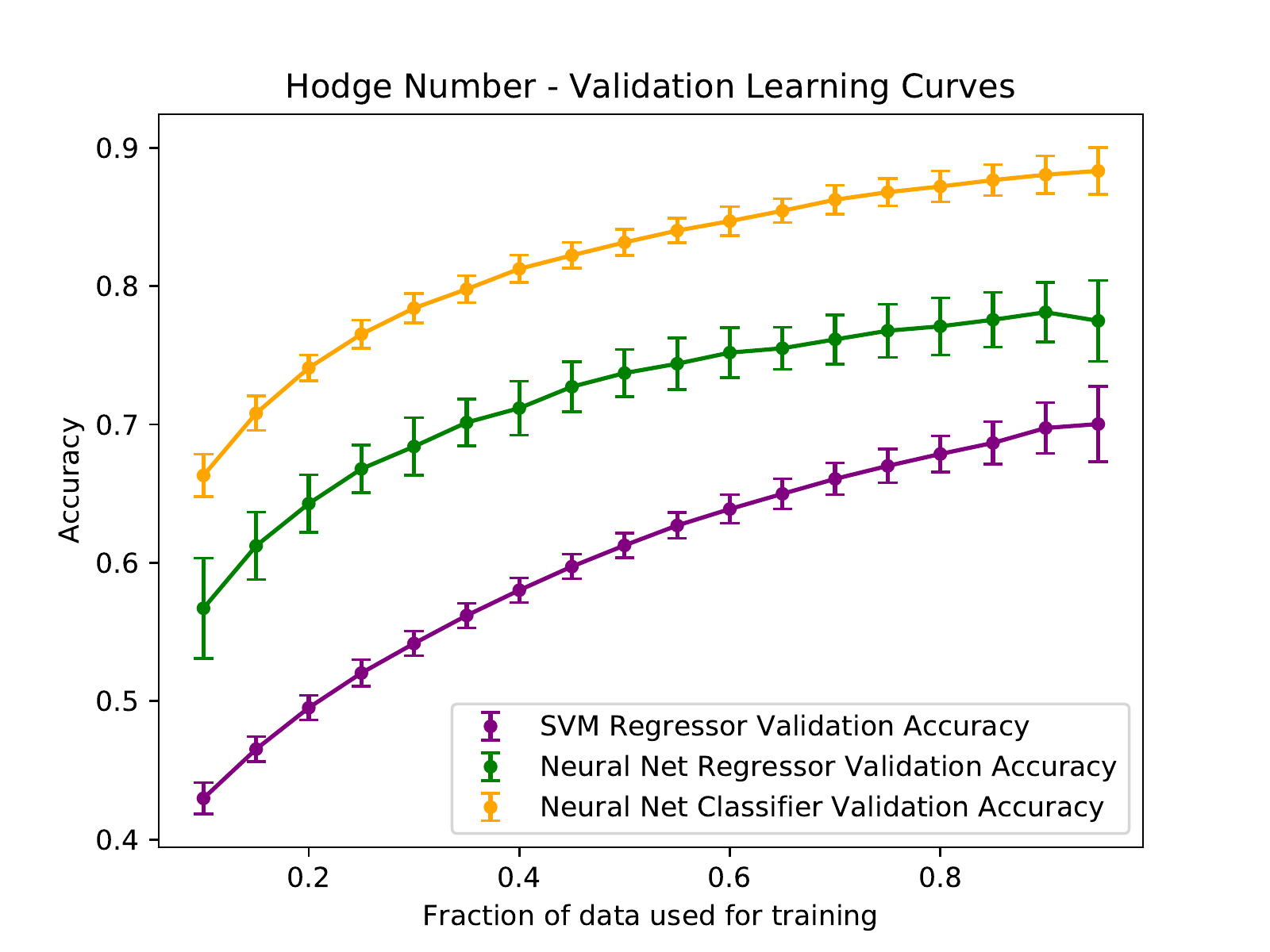}
	\vspace{5pt}
	\caption{{\sf Hodge learning curves generated by averaging over $100$ different
	random cross validation splits using a cluster. The accuracy quoted for the
	$20$ channel (since $h^{1,1} \in [0,19]$) neural network
	classifier is for complete agreement across all $20$ channels.}}
	\label{fig:hodge}
		\vspace{10pt}
\end{figure}
\begin{table}[h!!]
\centering
\footnotesize
\begin{tabular}{c|ccccc}
	& Accuracy & RMS & $R^2$ & WLB & WUB \\ \hline 
	SVM Reg & 0.70 $\pm$ 0.02 & \textbf{0.53}$\pm$ \textbf{0.06} & \textbf{0.78} $\pm$ \textbf{0.08} & 0.642 & 0.697  \\
	NN Reg & 0.78 $\pm$ 0.02 & 0.46 $\pm$ 0.05 & 0.72 $\pm$ 0.06 & 0.742 & 0.791 \\
	NN Class & \textbf{0.88} $\pm$ \textbf{0.02} & - & - & \textbf{0.847} & \textbf{0.886}\\
\end{tabular}
\vspace{5pt}
\caption{{\sf Summary of the highest validation accuracy achieved for predicting the Hodge numbers. WLB (WUB) stands for Wilson Upper (Lower) Bound. The dashes are because the NN classifer returns a binary 0/1 but RMS and $R^2$ are defined for continuous outputs.
We also include $99$\% Wilson confidence interval evaluated with a validation size of $0.25$ the total data ($1972$).
Errors were obtained by averaging over $100$ different random cross validation splits using a cluster.}}
\label{tab:hodge_acc} 
\end{table}

\begin{figure}[p]
	\centering
	\includegraphics[width=0.46\textwidth]{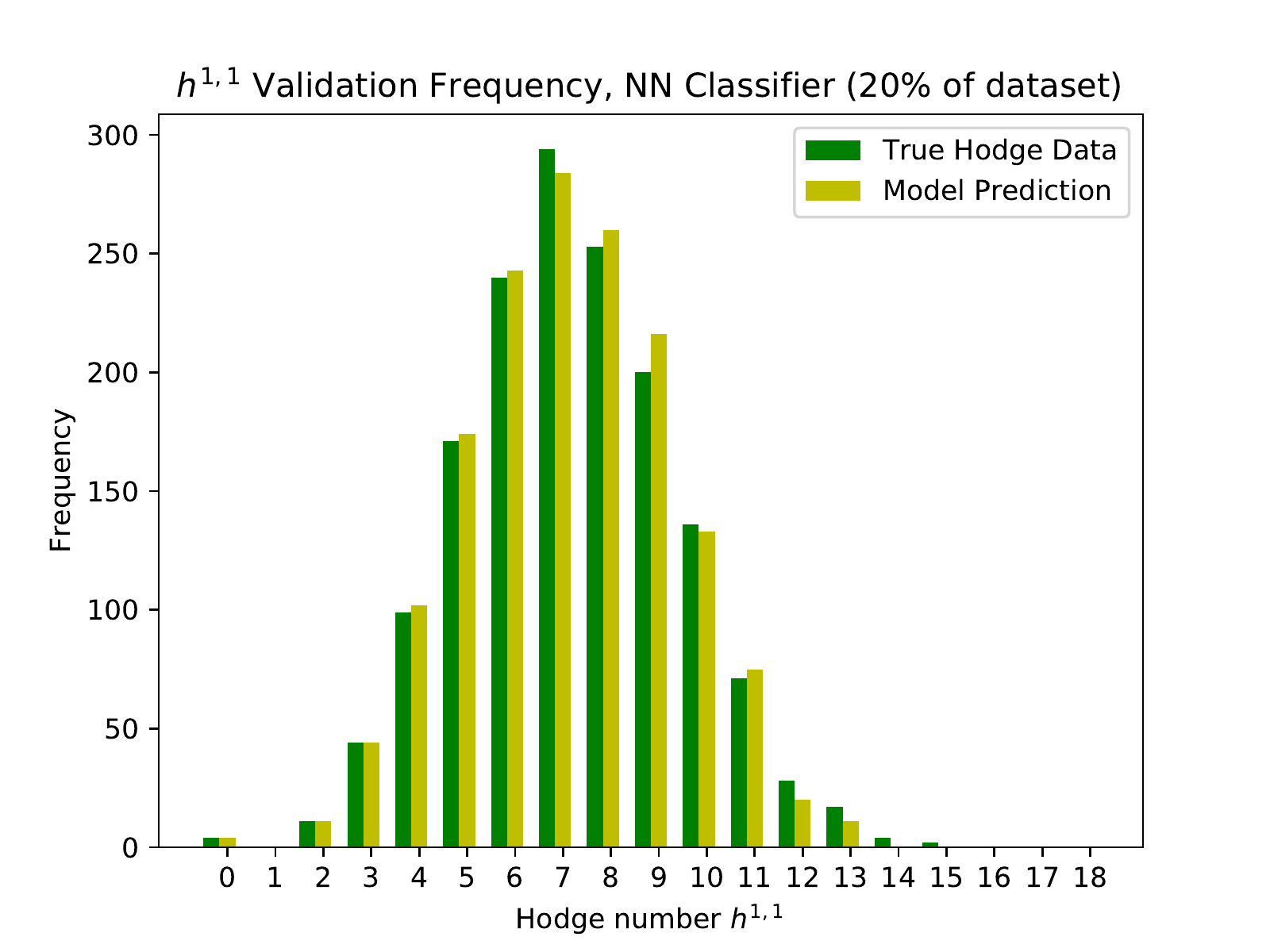}
	\includegraphics[width=0.46\textwidth]{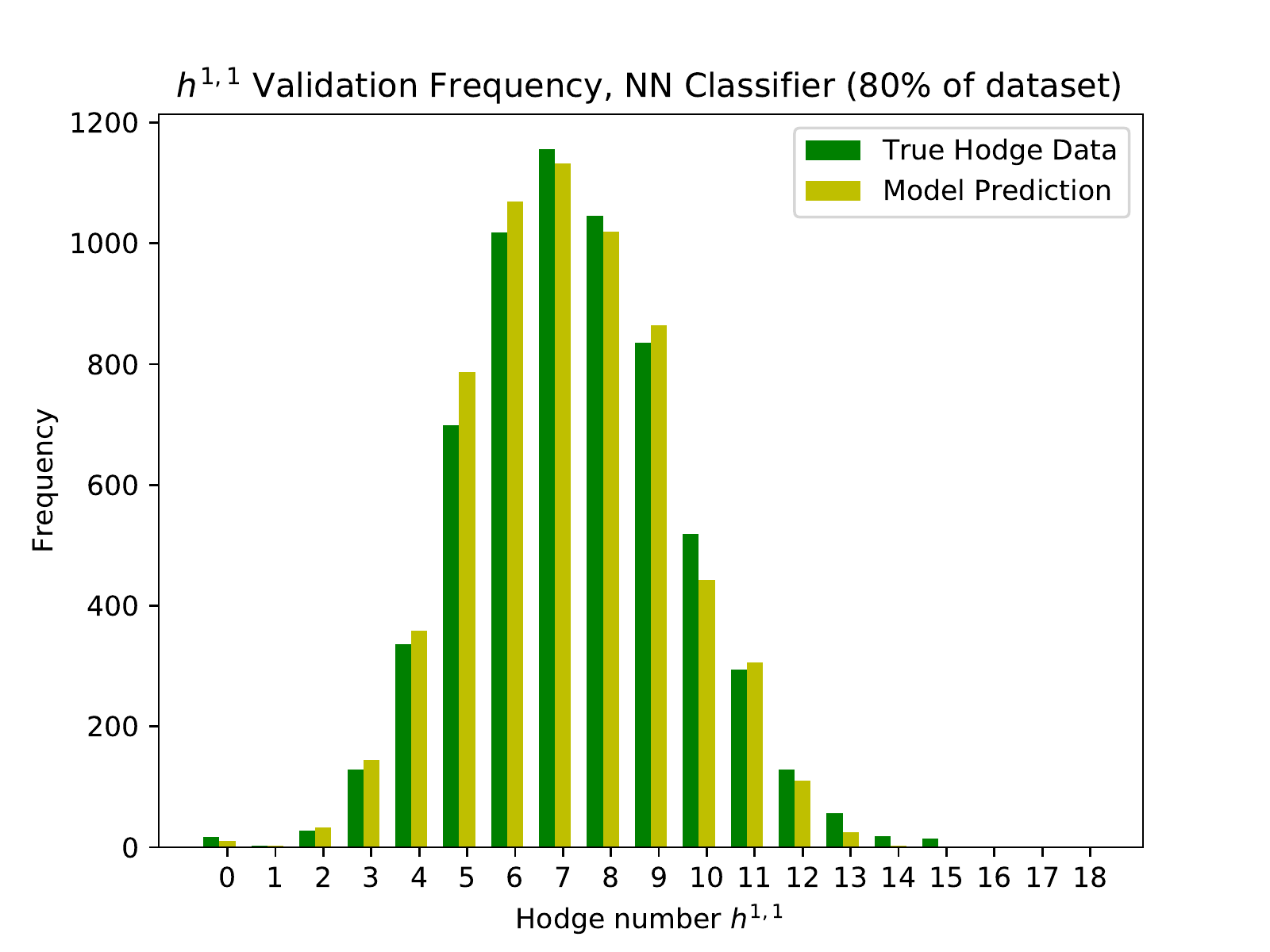}\\
	\vspace{10pt}
	\includegraphics[width=0.46\textwidth]{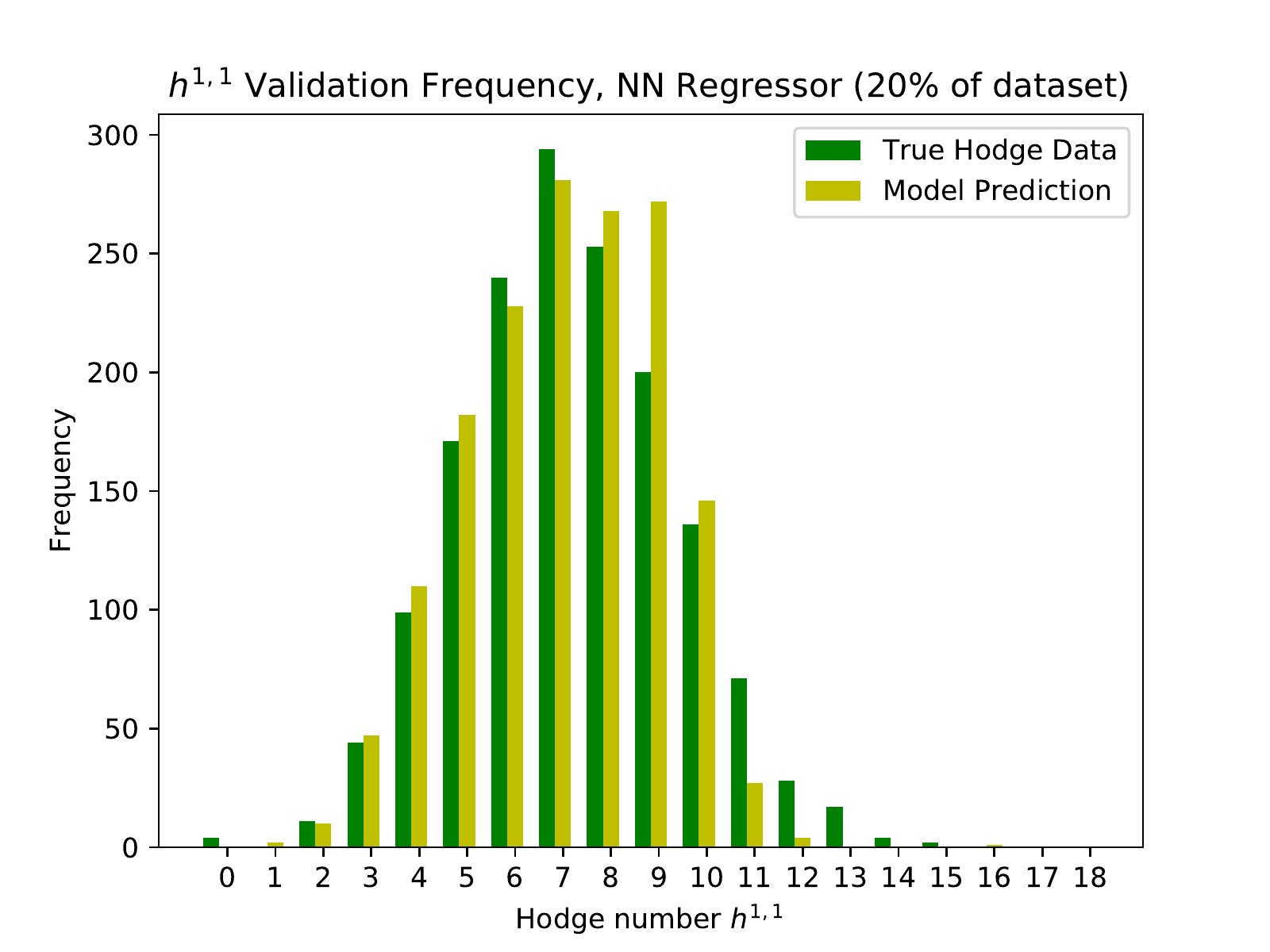}
	\includegraphics[width=0.46\textwidth]{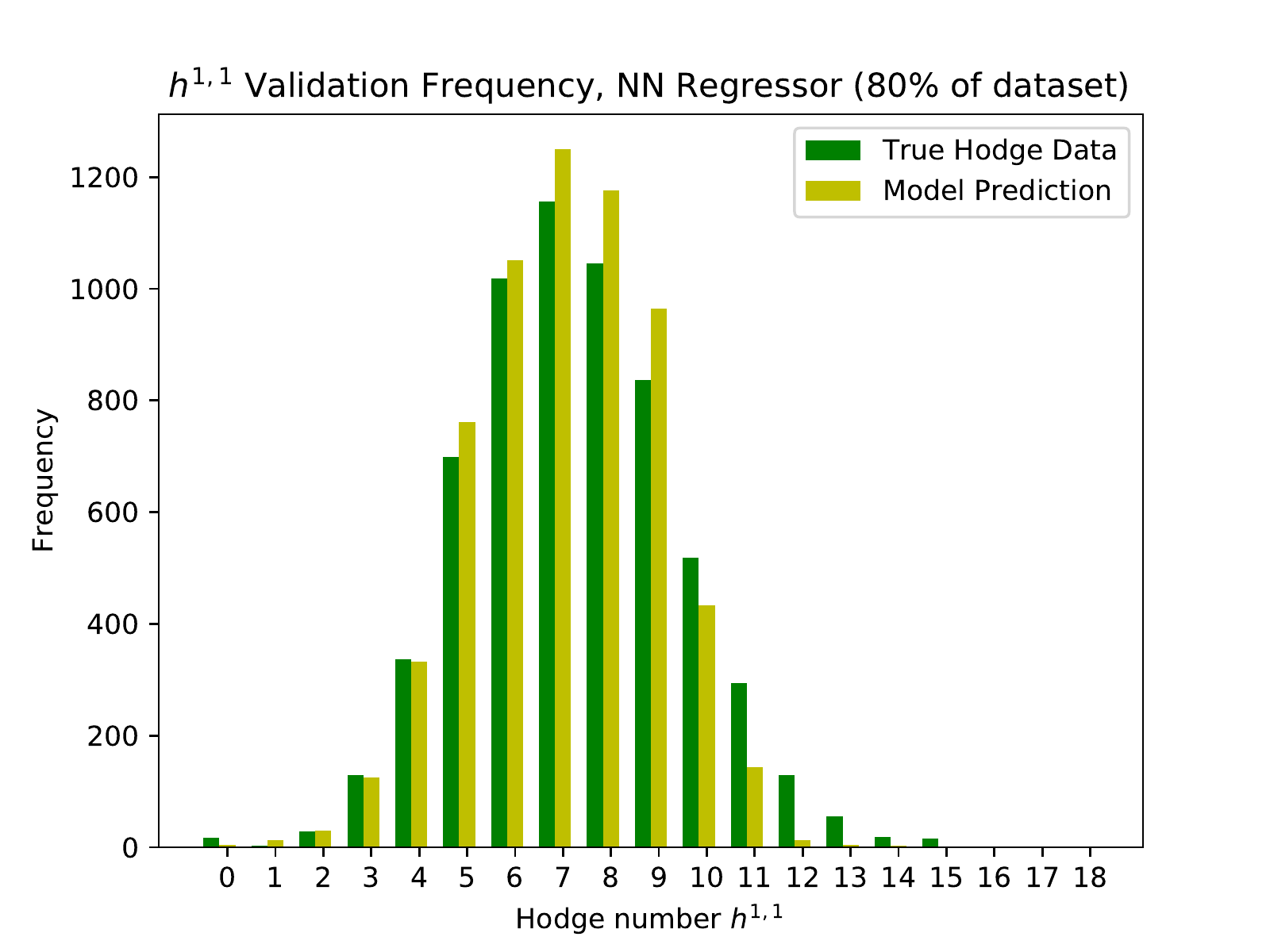}\\
	\vspace{10pt}
	\includegraphics[width=0.46\textwidth]{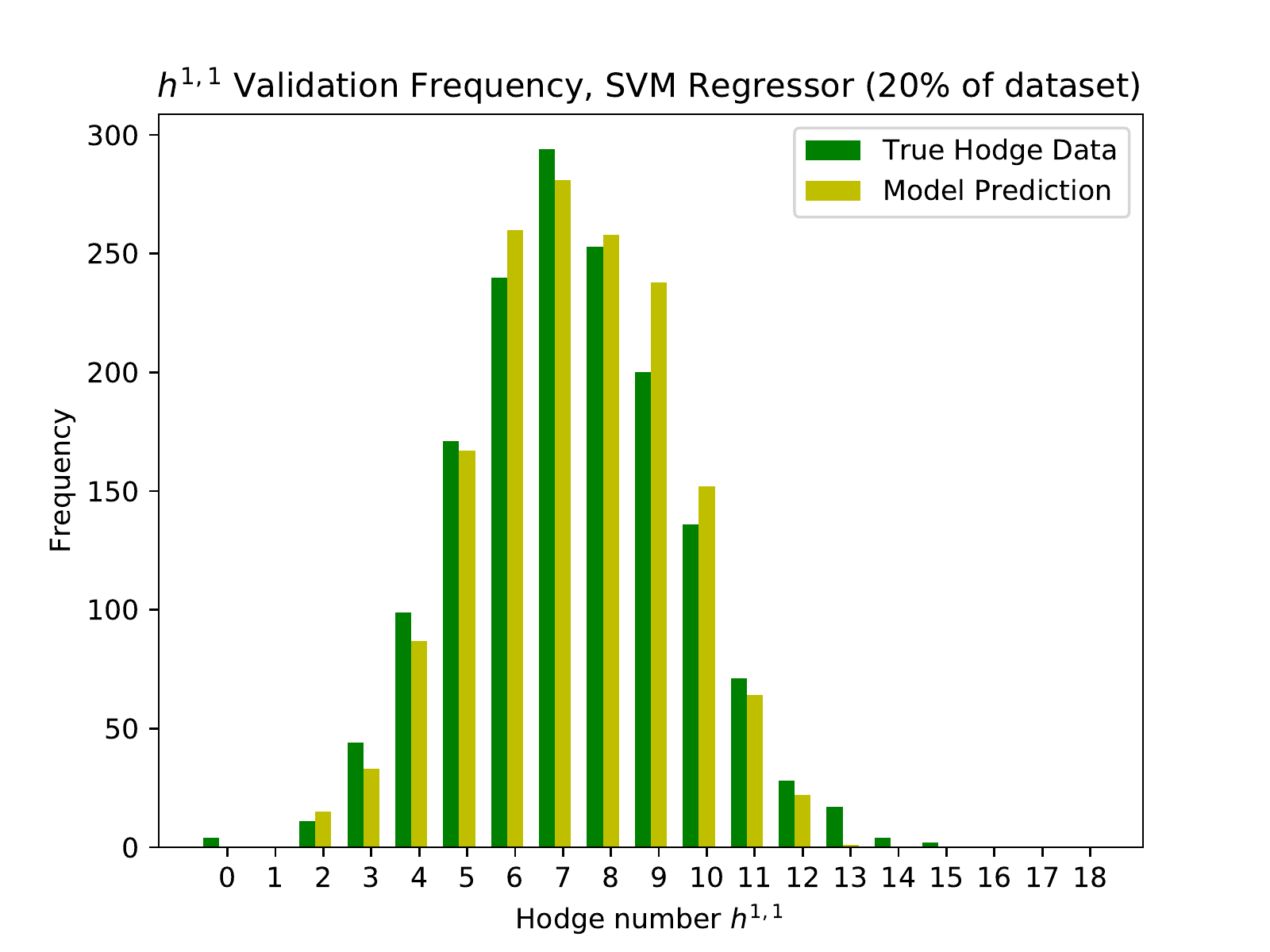}
	\includegraphics[width=0.46\textwidth]{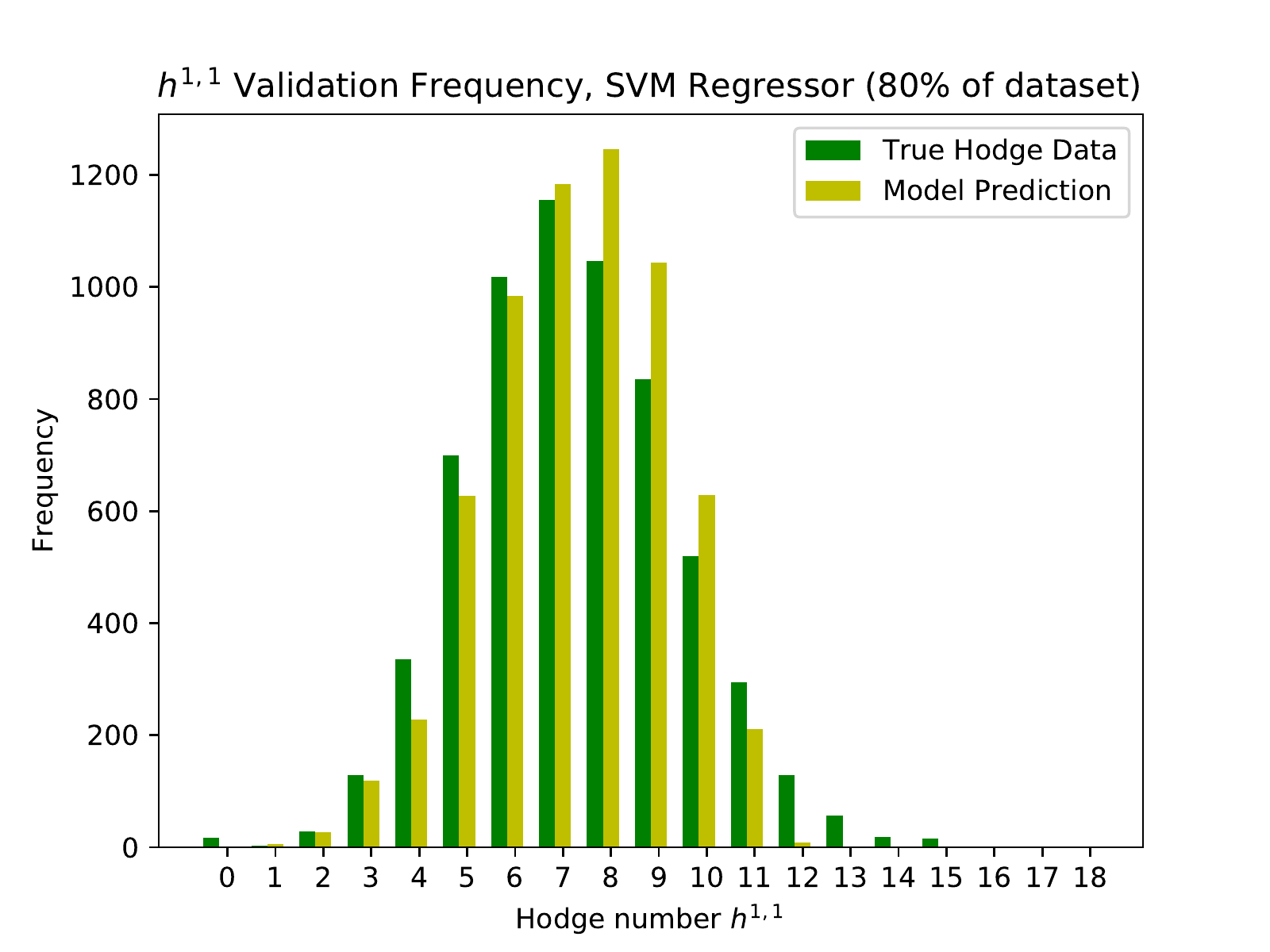}\\
	\vspace{5pt}
	\caption{{\sf The frequencies of $h^{1,1}$ (validation sets of size 20\% and 80\% respectively of the total data), for the neural network classifier (top row) and regressor (middle row) and the SVM regressor (bottom row).}}
	\label{fig:h11freqplotnn}
\end{figure}

\subsubsection{Outcomes}
Our results are summarised in Figures~\ref{fig:hodge} and \ref{fig:h11freqplotnn} and in \tref{tab:hodge_acc}.
Clearly, the validation accuracy improves as the training set increases in size.
The histograms in Figure \ref{fig:h11freqplotnn} show that the model slightly overpredicts at larger values of $h^{1,1}$.

We contrast our findings with the preliminary results of a previous case study by one of the authors \citep{he2017deep,He:2017set} in which a Mathematica implemented neural network of the multi-layer perceptron type was used to machine learn $h^{1,1}$.
In this work, a training data size of $0.63$ ($5000$) was used, and a test accuracy of $77\%$ was obtained.
Note this accuracy is against the entire dataset after seeing only the training set, whereas we compute validation
accuracies against only the unseen portion after training. 
In \citep{he2017deep} there were a total of $1808$ errors, so assuming the training set was perfectly learned (reasonable as training accuracy can be arbitrarily high with overfitting), this translates to a validation accuracy of $0.37$.
For the same sized cross validation split, we obtain a validation accuracy of $0.81 \pm 0.01$, a significant enhancement.
Moreover, it should be emphasized that whereas \citep{he2017deep,He:2017set} did a binary classification of large vs.~small Hodge numbers, here the {\it actual} Hodge number $h^{1,1}$ is learned, which is a much more sophisticated task.


\subsection{Machine Learning Favourable Embeddings} \label{fav}

\begin{figure}[th]
\centering
\includegraphics[width=0.67\textwidth]{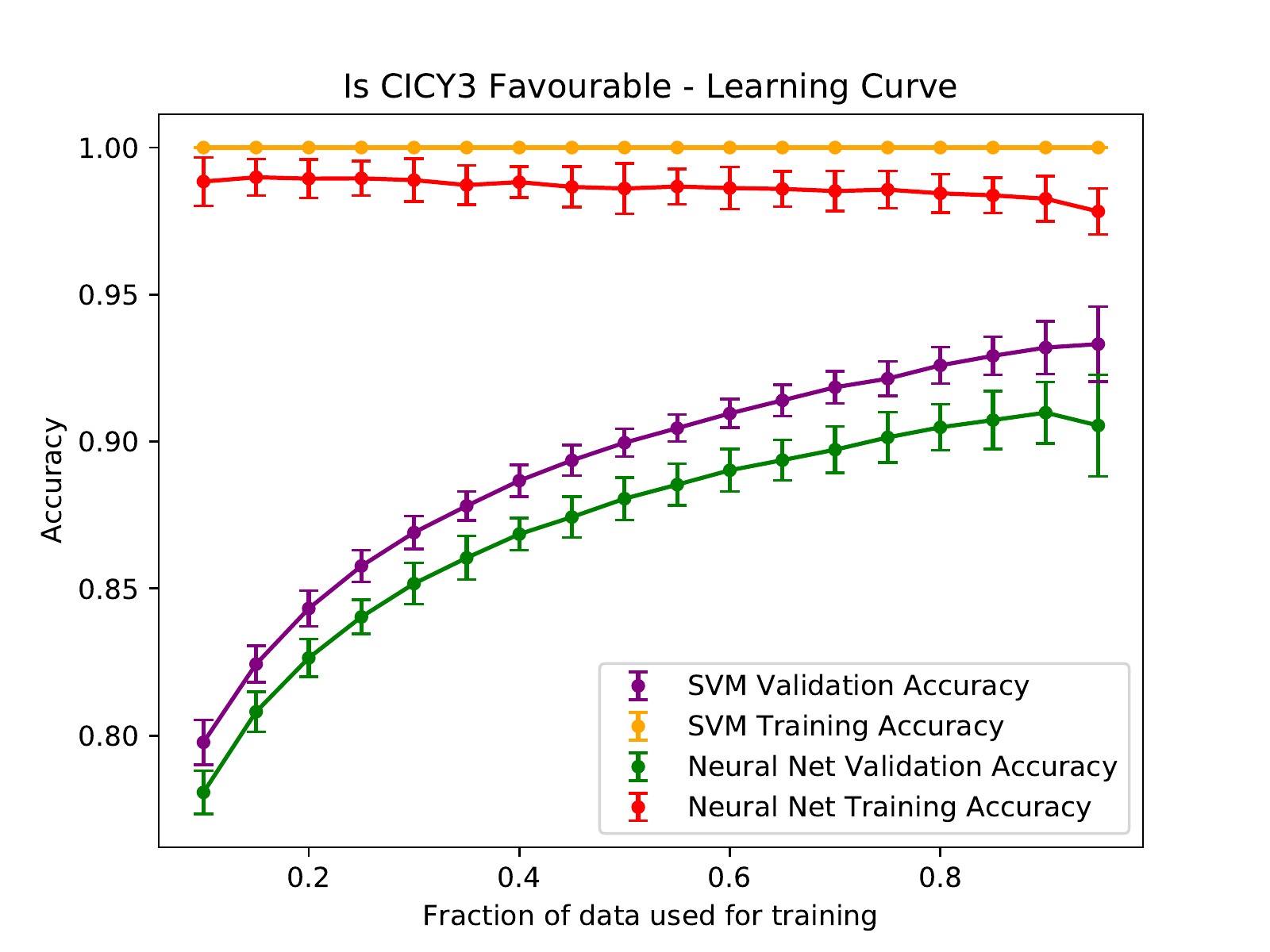}
\caption{{\sf Learning curves for testing favourability of a CICY.}}\label{fig:fav_LC}
\end{figure}

\begin{table}[h]
\centering
\small
\begin{tabular}{c|ccc}
	& Accuracy &  WLB & WUB \\ \hline
	SVM Class & \textbf{0.933} $\pm$ \textbf{0.013} & 0.867 &  0.893 \\
	NN Class & 0.905 $\pm$ 0.017 & \textbf{0.886} & \textbf{0.911}
\end{tabular} 
\vspace{5pt}
\caption{{\sf Summary of the best validation accuracy observed and $99\%$ Wilson confidence boundaries.
WLB (WUB) stands for Wilson Upper (Lower) Bound.
Errors were obtained by averaging over $100$ random cross validation splits using a cluster.}}\label{tab:fav_LC}
\end{table}

Following from the discussion in Section \ref{datasets}, we now study the binary query: given a CICY threefold configuration matrix \eqref{eq:confmatr1}, can we deduce if the CICY is favourably embedded in the product of projective spaces?
Already we could attempt to predict if a configuration is favourable with the results of Section \ref{hodge} by predicting $h^{1,1}$ explicitly and comparing it to the number of components of ${\IA}$.
However, we rephrase the problem as a binary query, taking the CICY configuration matrix as the input and return $0$ or $1$ as the output.

An optimal SVM architecture was found by hand to use a Gaussian kernel with $\sigma{=}3$ and $C{=}0$. Neural network architecture was also found by hand, as a simple one hidden layer neural network with $985$ neurons, ReLU activation, dropout of $0.46$, and sigmoid activation at the output layer gave best results.

Results are summarised in \fref{fig:fav_LC} and \tref{tab:fav_LC}.
Remarkably, after seeing only $5\%$ of the training data ($400$ entries), the models are capable of extrapolating to the full dataset with an accuracy $\sim 80\%$.
This analysis took less than a minute on a laptop computer.
Since computing the Hodge numbers directly was a time consuming and nontrivial problem \citep{bestiary}, this is a prime example of how applying machine learning could shortlist different configurations for further study in the hypothetical situation of an incomplete dataset.

\subsection{Machine Learning Discrete Symmetries} \label{sym}
The symmetry data resulting from the classifications \cite{Candelas:2008wb,
free_syms} presents various properties that we can try to machine learn. An
ideal machine learning model would be able to replicate the classification
algorithm, giving us a list of every symmetry group which is a quotient for a
given manifold. However, this is a highly imbalanced problem, as only a tiny
fraction of the $7890$ CICYs would admit a specific symmetry group. Thus, we
first try a more basic question, given a CICY configuration, can we predict if
the CICY admits any freely acting group. This is still most definitely a
\textit{needle in a haystack} problem as only $2.5\%$ of the data belongs to the
true class. In an effort to overcome this large class imbalance, we generate new
synthetic data belonging to the positive class. We try two separate methods to
achieve this - \textbf{sampling techniques} and \textbf{permutations of the CICY matrix}.

Sampling techniques preprocess the data to reduce the class imbalance.  For
example, downsampling drops entries randomly from the false class, increasing
the fraction of true entries at the cost of lost information.  Upsampling clones
entries from the true class to achieve the same effect. This is effectively the
same as associating a larger penalty (cost) to misclassifying entries in the
minority class. Here, we use Synthetic Minority Oversampling Technique (SMOTE)
\citep{smote} to boost performance.

\subsubsection{SMOTE}
SMOTE is similar to upsampling as it increases the entries in the minority class as opposed to downsampling.
However, rather than purely cloning entries, new \textit{synthetic} entries are created from the coordinates of entries in the feature space.
Thus the technique is ignorant to the actual input data and generalises to any machine learning problem.
We refer to different \text{amounts} of SMOTE by a integer multiple of $100$.
In this notation, SMOTE $100$ refers to doubling the minority class ($100\%$ increase), SMOTE $200$ refers to tripling the minority class and so on:
\subsubsection*{SMOTE Algorithm}
\vspace{5pt}
\begin{enumerate}
\item For each entry in the minority class $\mathbf{x_i}$, calculate its $k$ nearest neighbours $\mathbf{y_k}$ in the feature space (\textit{i.e.}, reshape the $12 \times 15$, zero padded CICY configuration matrix into a vector $\mathbf{x_i}$, and find the nearest neighbours in the resulting $180$ dimensional vector space).

\item Calculate the difference vectors $\mathbf{x_i - y_k}$ and rescale these by a random number $n_k \in (0,1)$.

\item Pick at random one point $\mathbf{x_i} + n_k (\mathbf{x_i - y_k})$ and keep this as a new synthetic point.

\item Repeat the above steps $N/100$ times for each entry, where $N$ is the amount of SMOTE desired.
\end{enumerate}

\subsubsection{SMOTE Threshold sweep, ROC, and $F$-values}\label{s:ROC}
\begin{figure}[h!!!]
	\centering
	\includegraphics[width=0.5\textwidth]{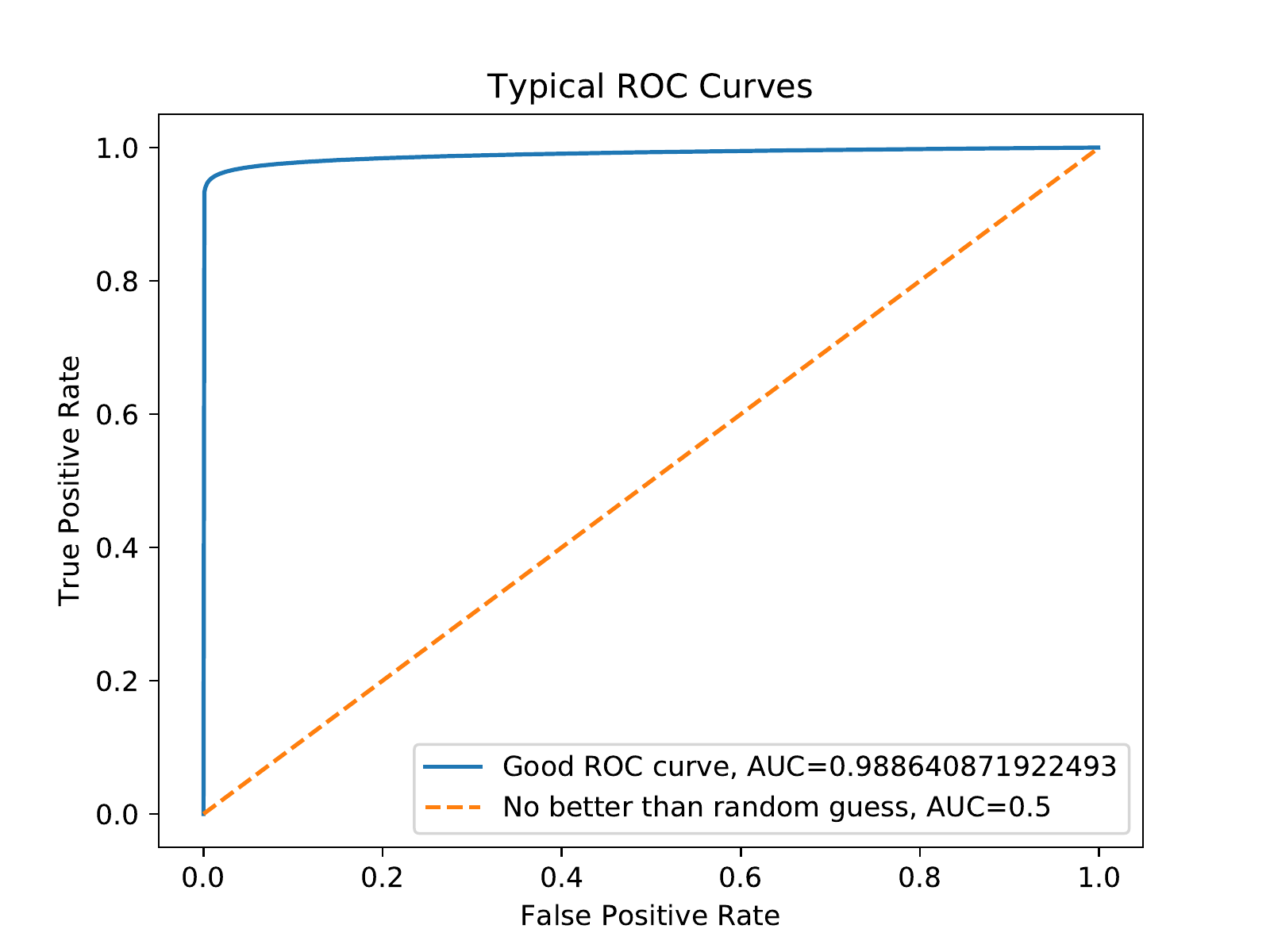}
	\vspace{5pt}
	\caption{{\sf Typical ROC curves.
	The points above the diagonal represent classification results which are better than random.}}
	\label{fig:typ_ROC}
		\vspace{10pt}
\end{figure}

The results obtained here all trivially obtained validation accuracies $\sim 99\%$.
As noted in Section~\ref{benchmark}, this is meaningless and instead we should use AUC and $F$-values as our metrics.
However, after processing the data with a sampling technique and training the model, we would only obtain one point $(\mathrm{FPR},\mathrm{TPR})$ to plot on a ROC curve.
Thus, to generate the full ROC curve, we vary the output threshold of the model to sweep through the entire range of values.
More explicitly, for an SVM, we modify the classifying function $\text{sgn} (f(\mathbf{x})) $ to $\text{sgn}(f(\mathbf{x})-t)$.
For neural networks, we modify the final sigmoid activation layer \eqref{eq:nn_sig} to be $\sigma_i^N = \sigma(W_{ij}^N \sigma_j^{N-1} + b_i^N - t)$.
Sweeping through all values of $t$ we generate the entire ROC curve and a range of $F$-values, thus obtaining the desired metrics.
Figure \ref{fig:typ_ROC} shows the profile of a good ROC curve.

\subsubsection{Permutations}
From the definition of the CICY \textit{configuration matrix}
\eqref{eq:confmatr1}, we note that row and column permutations of this matrix
will represent the same CICY. Thus we can reduce the class imbalance by simply
including these permutations in the training data set. In this paper we use the
same scheme for different amounts of PERM as we do for SMOTE, that is, PERM
$100$ doubles the entries in the minority class, thus one new permuted matrix
is generated for each entry belonging to the positive class.  PERM $200$ creates
two new permuted matrices for each entry in the positive class. Whether a row or
column permuation is used is decided randomly.

\subsubsection{Outcomes}
Optimal SVM hyperparameters were found by hand to be a Gaussian kernel with $\sigma=7.5,C=0$.
A genetic algorithm found the optimal neural network architecture to be three hidden layers with $287$, $503$, and $886$ neurons, with ReLU activations and a dropout of $0.4914$ in between each layer.

\begin{figure}[h]
	\centering
	\includegraphics[width=0.46\textwidth]{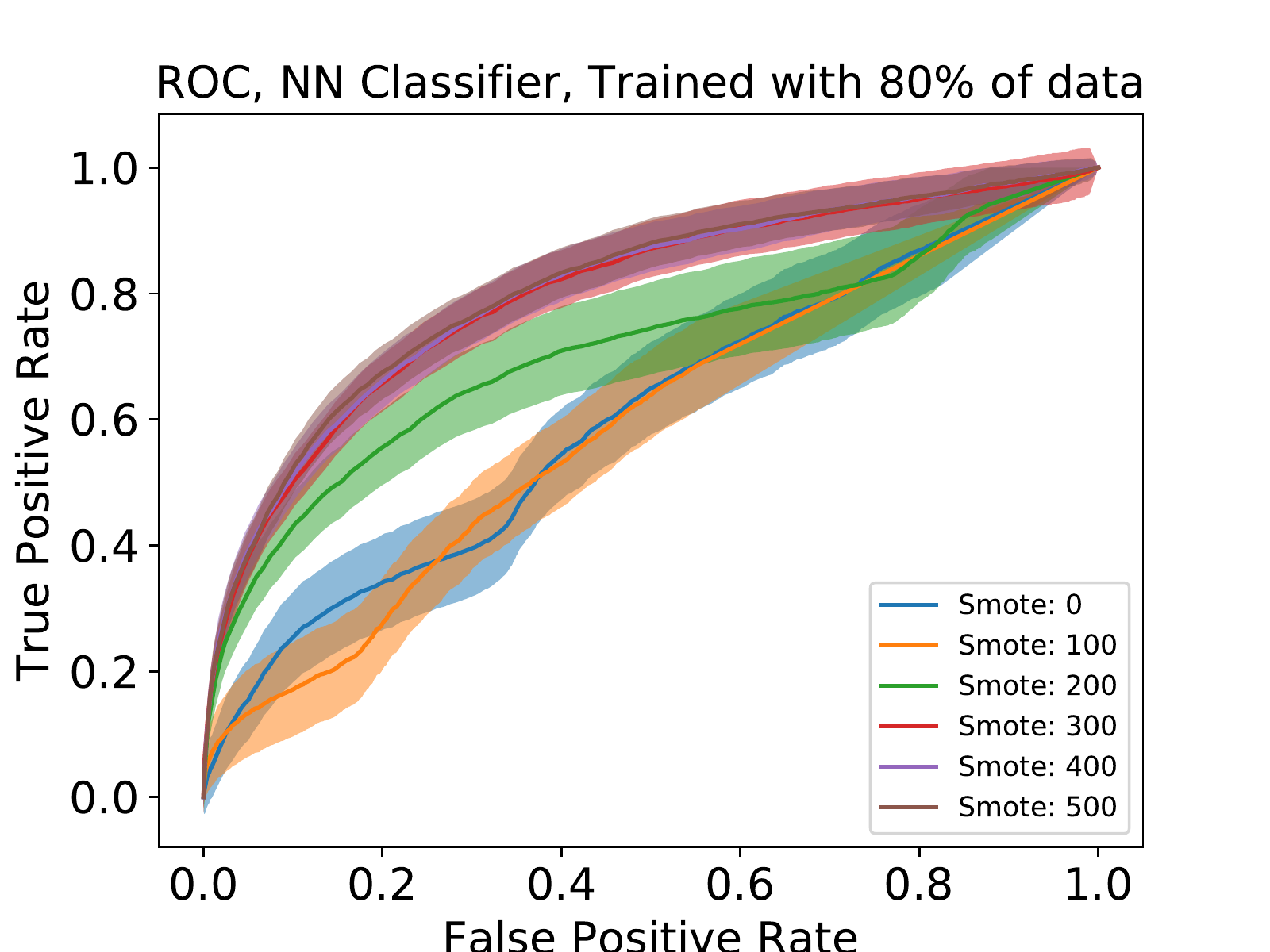}
	\includegraphics[width=0.46\textwidth]{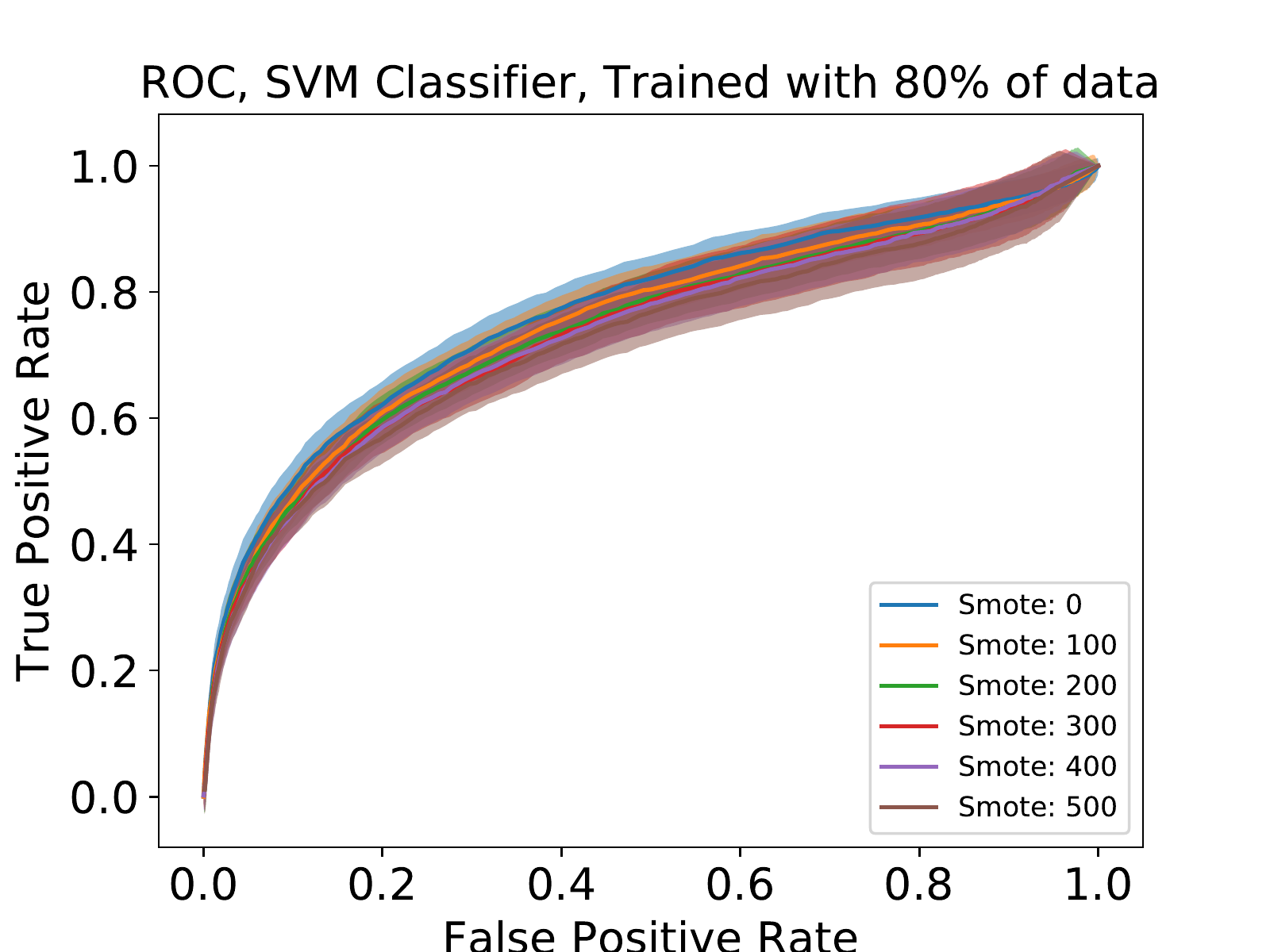}\\
	\vspace{10pt}
	\includegraphics[width=0.46\textwidth]{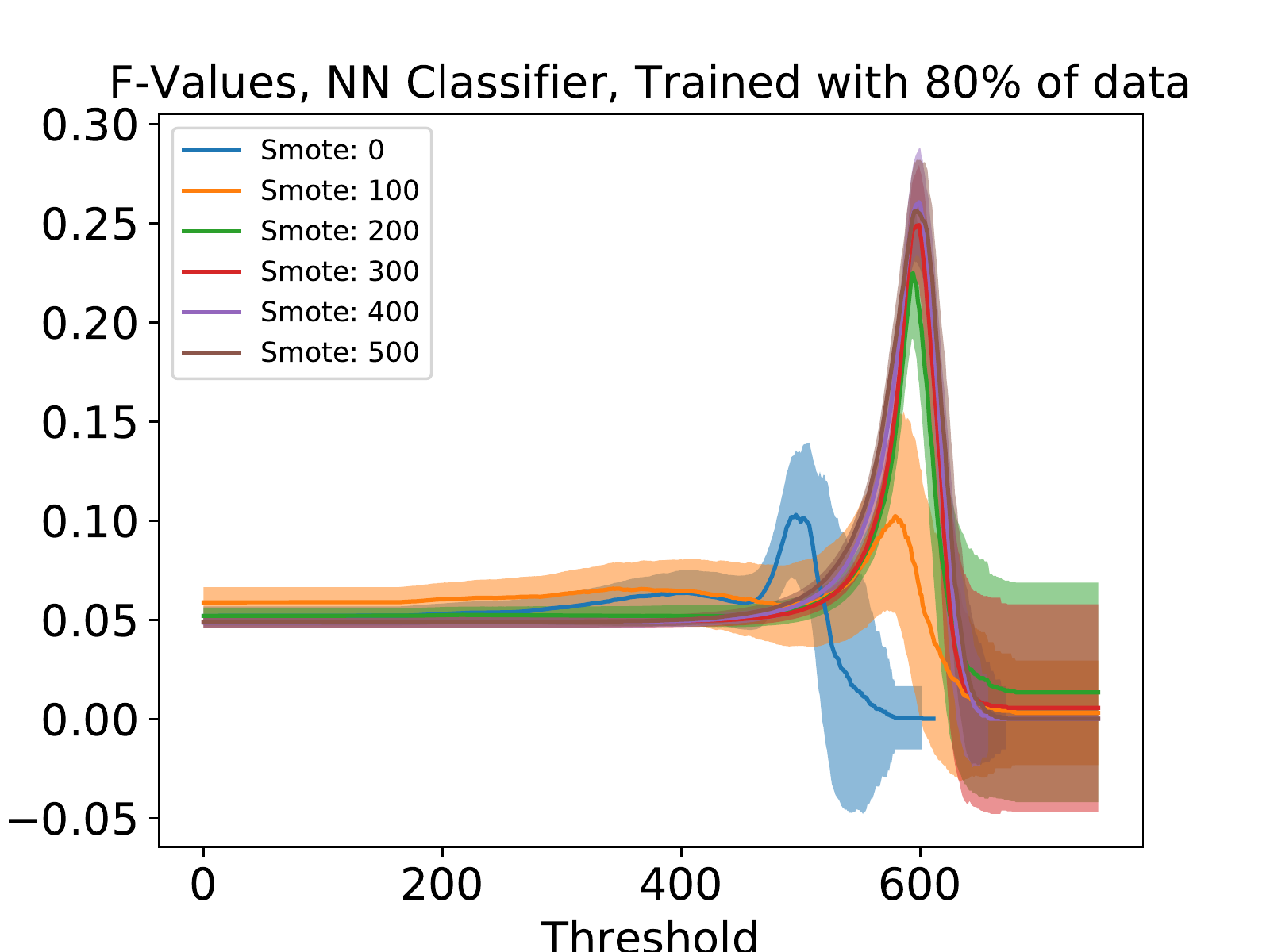}
	\includegraphics[width=0.46\textwidth]{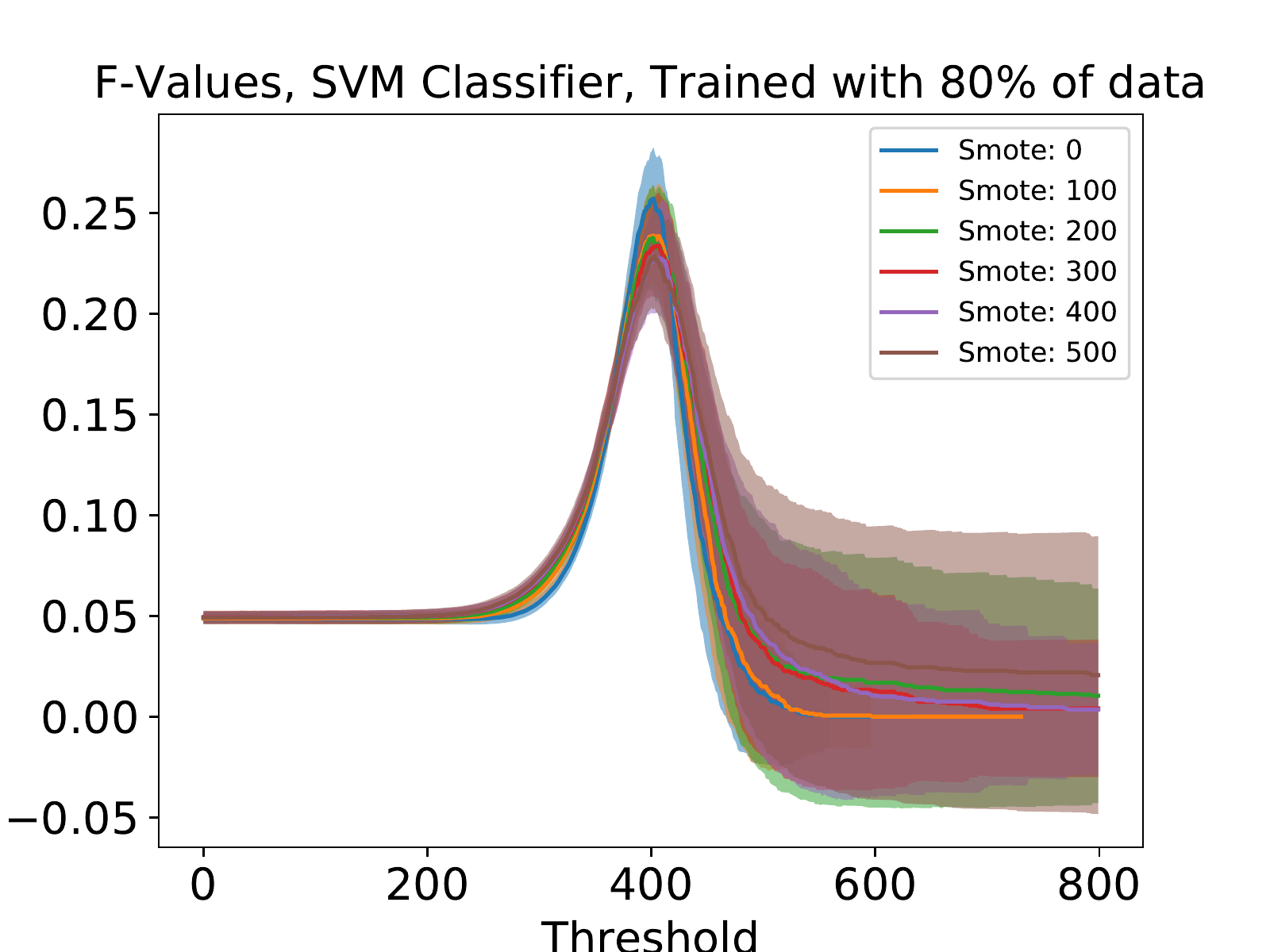}
	\vspace{5pt}
	\caption{{\sf ROC and $F$-curves generated for both SVM and neural network for several SMOTE values
	by sweeping thresholds and averaging over $100$ different random
cross validation splits. Here we present results after the models have been
trained on $80\%$ of the the training data. Shading represents possible values
to within one standard deviation of measurements.}}
	\label{fig:sym_graphs}
		\vspace{10pt}
\end{figure}

\begin{table}
\centering
\footnotesize
\begin{tabular}{c|c|c|c|c}
	{SMOTE} & SVM AUC & SVM max F & NN AUC & NN max F \\ \hline
	0 & 0.77 $\pm$ 0.03 & 0.26 $\pm$ 0.03 & 0.60 $\pm$ 0.05 & 0.10 $\pm$ 0.03 \\
	100 & 0.75 $\pm$ 0.03 & 0.24 $\pm$ 0.02 & 0.59 $\pm$ 0.04 & 0.10 $\pm$ 0.05 \\
	200 & 0.74 $\pm$ 0.03 & 0.24 $\pm$ 0.03 & 0.71 $\pm$ 0.05 & 0.22 $\pm$ 0.03 \\
	300 & 0.73 $\pm$ 0.04 & 0.23 $\pm$ 0.03 & 0.80 $\pm$ 0.03 & 0.25 $\pm$ 0.03 \\
	400 & 0.73 $\pm$ 0.03 & 0.23 $\pm$ 0.03 & 0.80 $\pm$ 0.03 & 0.26 $\pm$ 0.03 \\
	500 & 0.72 $\pm$ 0.04 & 0.23 $\pm$ 0.03 & 0.81 $\pm$ 0.03 & 0.26 $\pm$ 0.03 \\
\end{tabular}
	\vspace{5pt}
\caption{{\sf Metrics for predicting freely acting symmetries.
Errors were obtained by averaging over $100$ random cross validation splits using a cluster.}}
\label{tab:free_smote}
\end{table}

SMOTE results are summarised in Table \ref{tab:free_smote} and Figure \ref{fig:sym_graphs}.
As we sweep the output threshold, we sweep through the extremes of classifying everything as true or false, giving the ROC curve its characteristic shape.
This also explains the shapes of the $F$-value graphs.
For everything classified false, $tp,fp \rightarrow 0$, implying the $F$-value blows up, hence the diverging errors
on the right side of the $F$-curves.
For everything classified true, $fp\gg tp$ (as we only go up to SMOTE 500 with $195$ true entries and use $80\%$ of the training data, this approximation holds).
Using a Taylor expansion $F \approx 2 tp/fp = 2 \times 195/7890 = 0.049$. 
This is observed on the left side of the $F$-curves.
In the intermediate stage of sweeping, there will be an optimal ratio of true and false positives, leading to a maximum of the $F$-value.
We found that SMOTE did not affect the performance of the SVM.
Both $F$-value and ROC curves for various SMOTE values are all identical within one standard deviation.
As the cost variable for the SVM $C=0$ (ensuring training leads to a global minimum), this suggests that the synthetic entries are having no effect on the generated hypersurface.
The distribution of points in feature space is likely too strongly limiting the possible regions synthetic points can be generated.
However, SMOTE did lead to a slight performance boost with the neural network.
We see that SMOTEs larger than 300 lead to diminishing returns, and again  the results were quite poor, with the largest $F$-value obtained being only $0.26$.
To put this into perspective, the optimal confusion matrix values in one run for this particular model (NN, SMOTE 500) were
$tp=30,tn=1127,fp=417,fn=10$.
Indeed, this model could be used to shortlist $447$ out of the $1584$ for further study, but $417$ of them are falsely predicted to have a symmetry and worse still this model misses a quarter of the actual CICYs with a symmetry.

\begin{figure}[h]
	\centering
	\includegraphics[width=0.55\textwidth]{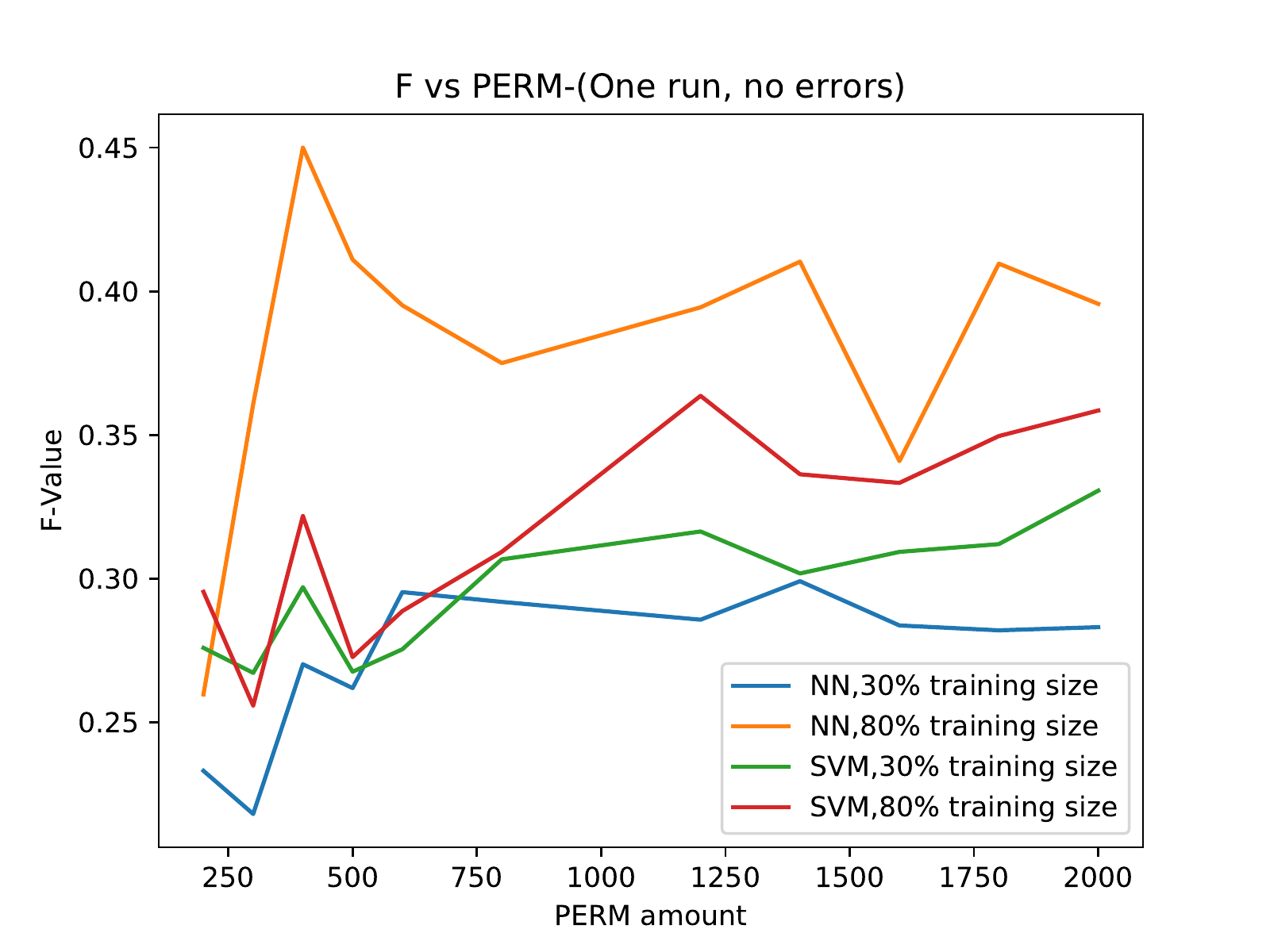}
	\caption{{\sf Plot of permutation F-values up to PERM $2000$}}
	\label{fig:free_perm_graph}
		\vspace{10pt}
\end{figure}

PERM results are summarized in Table \ref{tab:free_perm} and Figure
\ref{fig:free_perm_graph}. Note these results are not averaged over several runs
and are thus noisy. We see that for $80\%$ of the training data used (the same
training size as used for SMOTE runs) that the F-values are of the order
$0.3-0.4$. This is a slight improvement over SMOTE but we note from the PERM
$100,000$ results in Table \ref{tab:free_perm} there is a limit to the improvement
permutations can give.

Identifying the existence and the form of freely acting discrete symmetries on a Calabi--Yau geometry is a difficult mathematical problem.
It is therefore unsurprising that the machine learning algorithms also struggle when confronted with the challenge of finding a rare feature in the dataset.

\begin{table}
\centering
\footnotesize
\begin{tabular}{c|c|c|c|c}
	& \multicolumn{2}{c|}{{30\% Training Data}} & \multicolumn{2}{c}{{80\%
	Training Data}} \\ \hline
	PERM & NN F-Value & SVM F-Value & NN F-Value & SVM F-Value \\
	100000 & 0.2857 & -- & 0.3453 & -- \\
	10000 & 0.3034 & 0.2989 & 0.3488 & -- \\
	2000 & 0.2831 & 0.3306 & 0.3956 & 0.3585 \\
	1800 & 0.2820 & 0.3120 & 0.4096 & 0.3486 \\
	1600 & 0.2837 & 0.3093 & 0.3409 & 0.3333 \\
	1400 & 0.2881 & 0.3018 & 0.4103 & 0.3364 \\
	1200 & 0.2857 & 0.3164 & 0.3944 & 0.3636 \\
	800 & 0.2919 & 0.3067 & 0.3750 & 0.3093 \\
	600 & 0.2953 & 0.2754 & 0.3951 & 0.2887 \\
	500 & 0.2619 & 0.2676 & 0.4110 & 0.2727 \\
	400 & 0.2702 & 0.2970 & \textbf{0.4500} & 0.3218 \\
	300 & 0.2181 & 0.2672 & 0.3607 & 0.2558 \\
	200 & 0.2331 & 0.2759 & 0.2597 & 0.2954 \\
\end{tabular}
	\vspace{5pt}
\caption{{\sf F-Values obtained for different amounts of PERMS for one run.
	Dashes correspond to experiments which couldn't be run due to memory errors.}}
\label{tab:free_perm}
	\vspace{10pt}
\end{table}

\section{Discussion}
In this study, continuing with the paradigm in and improving upon the results of \cite{he2017deep, He:2017set},
we utilise neural networks and Support Vector Machines (SVMs) to machine learn various geometric  properties of CICY threefolds.
Neural networks were implemented using the {Keras} Python package with {TensorFlow} backend.
SVMs were implemented by using the quadratic programming Python package {Cvxopt} to solve the SVM optimisation
problem.
To benchmark the performance of each model we use {cross validation} and take a variety of statistical measures where appropriate, including accuracy, Wilson confidence interval, $F$-values, and the area under the receiving operator characteristic (ROC) curve (AUC).
Errors were obtained by averaging over a large sample of cross validation splits and taking standard
deviations.
Models are optimised by maximising the appropriate statistical measure.
This is achieved either by varying the model by hand or by implementing a genetic algorithm. 
Remarkable accuracies can be achieved, even when, for instance, trying to predict the exact values of Hodge numbers.

This work serves as a proof of concept for exploring the geometric features of
Calabi--Yau manifolds using machine learning beyond binary classifiers and feedforward neural networks.
In future work, we intend to apply the same techniques to study the Kreuzer--Skarke \cite{ks} list of half a billion reflexive polytopes and the toric Calabi--Yau threefolds obtained from this dataset \cite{us}.
Work in progress extends the investigations in this paper to the CICY fourfolds \cite{CICY4} and cohomology of bundles over CICY threefolds.

\section*{Acknowledgements}
This paper is based on the Masters of Physics project of KB with YHH, both of whom would like to thank
the Rudolf Peierls Centre for Theoretical Physics, University of Oxford for the provision of resources.
YHH would also like to thank the Science and Technology Facilities Council, UK, for grant ST/J00037X/1, the Chinese
Ministry of Education, for a Chang-Jiang Chair Professorship
at NanKai University and the City of Tian-Jin for a Qian-Ren Scholarship, as well as Merton College, Oxford, for her enduring support.
VJ is supported by the South African Research Chairs Initiative of the DST/NRF.
He thanks Tsinghua University, Beijing, for generous hospitality during the completion of this work.
We thank participants at the ``Tsinghua Workshop on Machine Learning in Geometry and Physics 2018'' for comments.

\appendix
\section{Brief Overview of Neural Networks}\label{NeuralNetworks}
Neural networks are one branch of machine learning techniques, capable of
dealing with both classification and regression problems. There are several
different types of neural networks, but they all act as a non-trivial function
$f(\mathbf{v_{in}}) = \mathbf{v_{out}}$. We proceed to discuss feedforward neural
networks.

\subsection*{Feedforward Neural Networks} 
\begin{figure}[ht]
\begin{center}
{\includegraphics[width=0.5\textwidth]{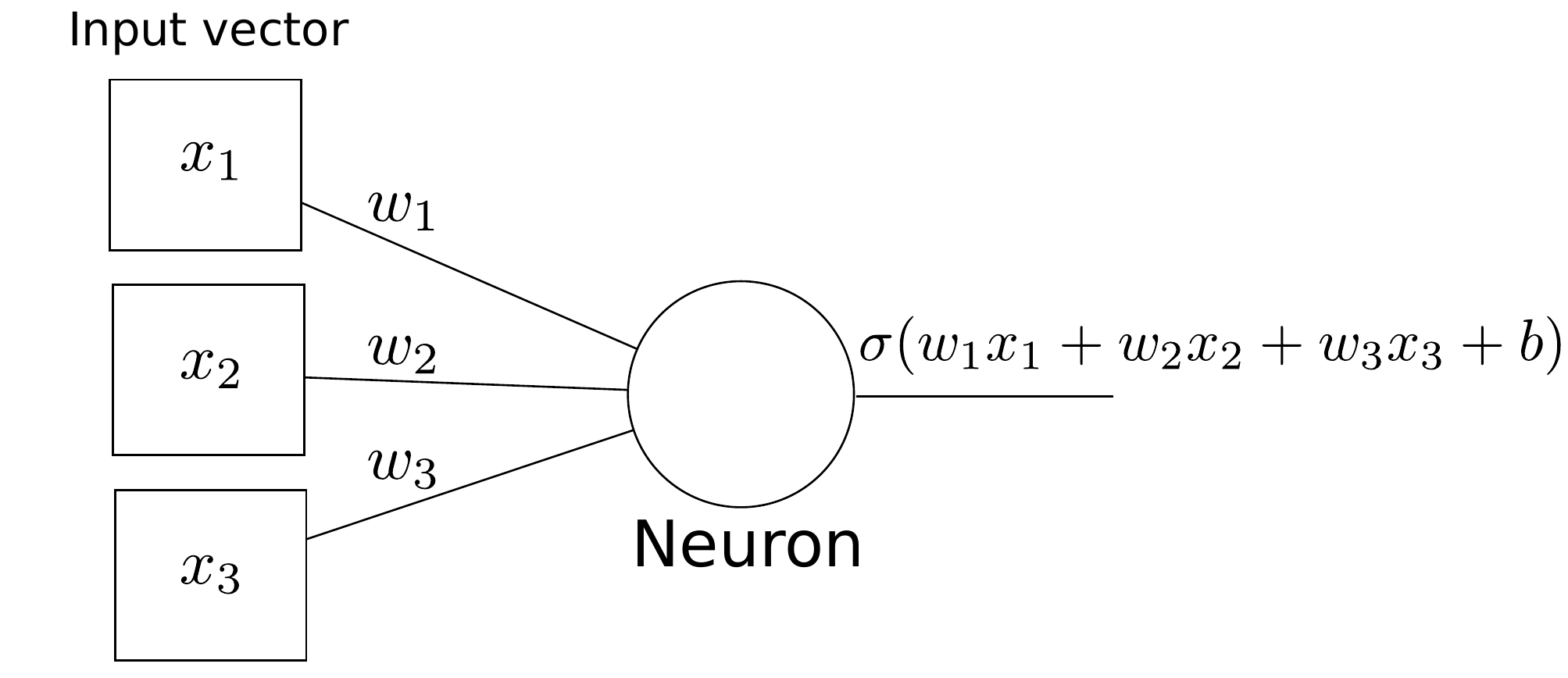}} \\[15pt]
\includegraphics[width=0.4\textwidth]{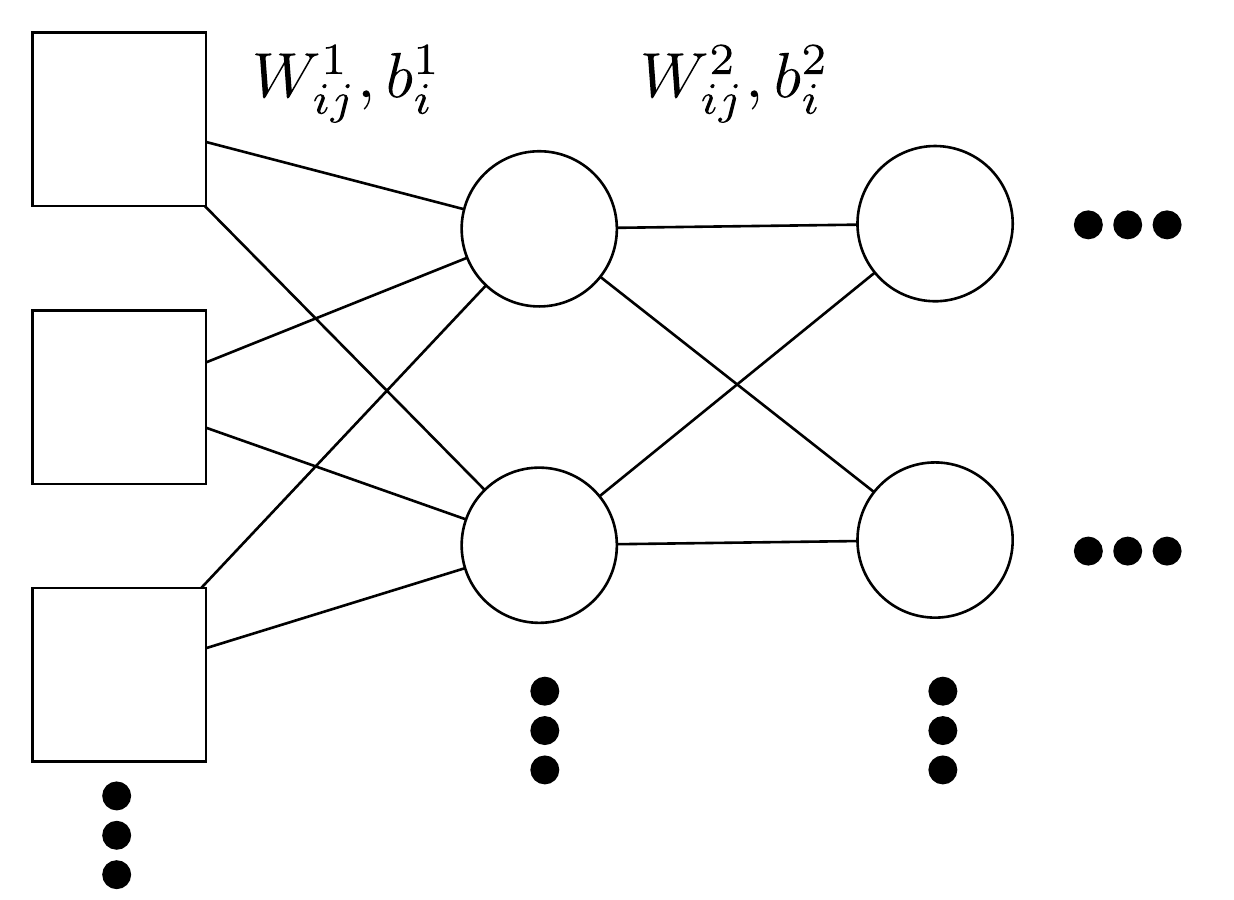}
\end{center}
	\vspace{5pt}
\caption{{\sf Schematic representation of feedforward neural network. The top figure denotes the perceptron (a single neuron), the bottom, the multiple neurons and multiple layers of the neural network.}}
\label{fig:nn_schem}
	\vspace{10pt}
\end{figure}
We first consider a single \textit{neuron}, which for an input $\mathbf{x}$ outputs
$\sigma(\mathbf{x \cdot w}+b)$.
Here $\sigma$ is the \textit{activation function}, $\mathbf{w}$ the \textit{weights} (for a weighted sum of the inputs), and $b$ the \textit{bias}. 

Activation functions are applied to the resulting sum, typically mapping to the
interval $[0,1]$. This mimics a real neuron, which is either {firing} or
not. The neuron can then act as a classifier (true or false) or a regressor
(with a suitable linear rescaling). Typical activation functions include the
sigmoid $\sigma(x) := 1 / (1+e^{-x})$, the rectified linear unit ReLU$(x):=\max(0,\alpha
x)$ and the function $\tanh$. 

A bias is included to offset the resulting weighted sum so as to stay in the \textit{active
region} of the activation function. To be more explicit, consider the sigmoid
activation. If we have a large input vector, without a bias, applying a sigmoid
activation function will tend to map the output to $1$ due to the large number
of entries in the sum, which may not be the correct response we expect from the
neuron. We could just decrease the weights, but training the net can
stagnate without a bias due to the vanishing gradient near $\sigma(x)=1,0$. 

We generalise to multiple neurons by promoting the weight vector to a
\textit{weight matrix}. This collection of neurons is known as a
\textit{layer}. We denote the output of the $i^\text{th}$ neuron in this layer as
$\sigma_i$
\begin{equation}
	\sigma_i := \sigma(W_{ij} x_j + b_i)\,.
\end{equation}
Extending to several layers, we just take the output of the previous layers as
the input to the next layer, applying a different weight matrix and bias vector
as we propagate through. Note that all internal layers between the input and
output layers are referred to as {hidden layers}. Denote the output of
the $i^\text{th}$ neuron in the $n^\text{th}$ layer as $\sigma^n_i$, with $\sigma^0_i = x_i$
as the input vector.
\begin{align}
	\label{eq:nn_sig}
	\sigma_i^n &= \sigma(W_{ij}^n \sigma_j^{n-1} + b_i^n)\,.
\end{align} 
This concludes how a fully connected neural network generates its output from a
given input. Training a network to give the desired output thus consists of
adjusting the weight and bias values. This is achieved by the back propagation
algorithm. 

\subsection*{Back Propagation in Neural Networks} \label{Back_prop}
To decide how to adjust the weights and biases of a neural network when
training, we define a cost function. Standard cost functions include mean
squared error and cross-entropy (categorical and binary). 
Back propagation is an algorithm to find parameter adjustments by minimising the
cost function. It is so named as adjustments are first made to the last layer
and then successive layers moving backwards.

To illustrate the approach consider a network with $M$ layers and a mean squared
error cost function
\begin{align}
E := \frac{1}{N} \sum_{train}^N \left( \bm{\sigma}^M-\mathbf{t}
\right)^2.
\end{align}
Here $N$ is the number of training entries and $\mathbf{t}$ the expected output for a
given entry.
Taking derivatives, shifts in the last weight matrix become:
\begin{align}
	\frac{\partial E}{\partial W_{ij}^M} &=\frac{2}{N} \sum_{train} (\sigma_i^M - t_i) \ \sigma_i^{'M} \sigma_j^{M-1}.
\end{align}
Working backwards, shifts in the second to last weight matrix:
\begin{align}
	\frac{\partial E}{\partial W_{ij}^{M-1}} &=\frac{2}{N} \sum_{train}\sum_u\  (\sigma_u^M - t_u)\  \sigma_u^{'M} W_{ui}^{M} \sigma_i^{' M-1}
 \sigma_j^{M-1} .
\end{align}
We define
\begin{align}
	\Delta_i^M &:= (\sigma_i^M - t_i) \sigma_i^{'M} , \quad\quad
	\Delta_i^m := \sum_u \Delta_u^{m+1} W_{ui}^{m+1} \sigma_i^{'m} .
\end{align}
Therefore by induction we can write (for an arbitrary layer $m$)
\begin{align}
	\frac{\partial E}{\partial W_{ij}^m} &= \frac{2}{N} \sum_{train}
	\Delta_i^m \sigma_j^{m-1}  , \quad\quad\frac{\partial E}{\partial b_{i}^m} = \frac{2}{N} \sum_{train}
	\Delta_i^m.
\end{align}
This defines the back propagation approach. By utilising our neural network's final output
and the expected output, we can calculate the $\Delta$s successively, starting from
the last layer and working backwards. We shift the weight values in the
direction the gradient is descending to minimise the error function. Thus shifts
are given by
\begin{align}
	\Delta W_{i,j}^m = - \eta \frac{\partial E}{\partial W_{ij}^m}\,,\quad \quad\Delta
	b_i^m = - \eta \frac{\partial E}{\partial b_i^m}\,.
\end{align}
With $\eta$ the {learning rate} (effectively a proportionality constant
fixing the magnitude of shifts in gradient descent). Care must be taken when
choosing the learning rate. A rate too small leads to slow convergence and the
possibility of becoming trapped in a local minimum. A rate too large leads to fluctuations
in errors and poor convergence as the steps taken in parameter space are too
large, effectively jumping over minima.

Note that parameter shifts are dependent on the gradient of the activation
function. For activation functions such as sigmoid or $\tanh$ this then drives the
output of a neuron to its minimal or maximal value, as parameter shifts become
increasingly small due to the vanishing gradient. This is advantageous in an
output layer where we may want to use binary classification. However, if neurons
in hidden layers are driven to their min/max too early in training, it can
effectively make them useless as their weights will not shift with any further
training. This is known as the flat spot problem and is why the ReLU activation
function has become increasingly popular.

\subsection*{Convolution Neural Networks} \label{conv} 
\begin{figure}[ht]
	\centering
	\includegraphics[width=0.4\textwidth]{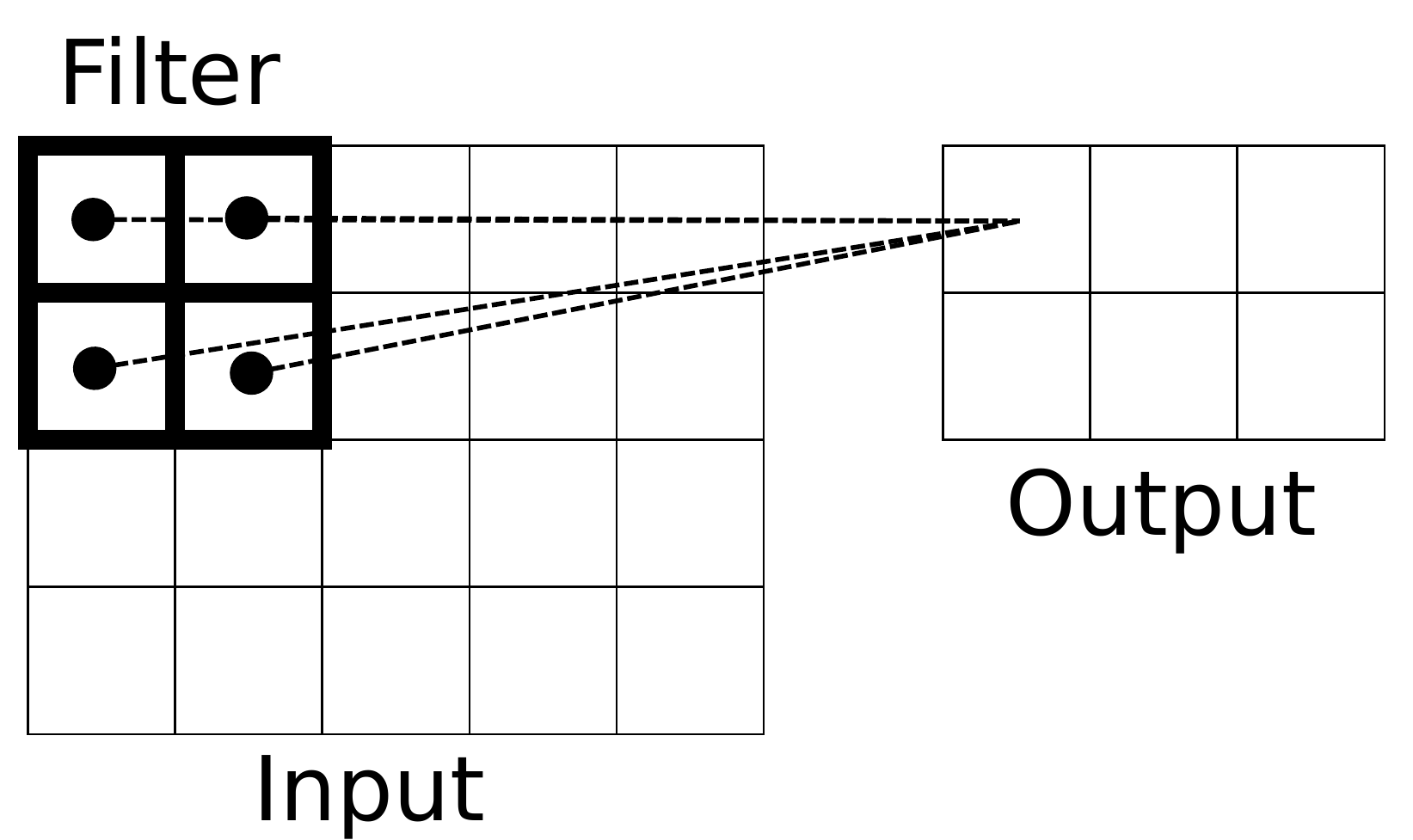}
	\vspace{5pt}
	\caption{{\sf Schematic of a convolution layer. Here the kernel has size $2
	\times 2$.  The neurons connected to each kernel window follow the window
systematically.  For example, when the kernel window moves one unit to the
right, the connected neuron in the output layer will be the centre square in the
top row.}}
	\label{fig:conv}
	\vspace{10pt}
\end{figure}

Convolutional neural networks (CNNs) are an alternative type of network which thrive when inputs contain
translational or rotational invariance. They are thus particularly useful for
image recognition. Just like a fully connected, feedforward layer, convolution
layers use a set of neurons which pass a weighted sum through an activation
function. However, neurons in convolution layers do not receive a weighted sum
from all the neurons in the previous layer. Instead, a {kernel} restricts
the contributing neurons (note a kernel in the context of CNNs is different to
SVMs). To be more explicit, consider a two dimensional input (matrix). A
{kernel} will be a grid sized $n \times n$ (arbitrary).  This grid
convolves across the input matrix, taking the smaller matrix the grid is
covering as the input for a neuron in the convolution layer, as shown in Figure
\ref{fig:conv}.
The output generated by convolving the kernel across the input matrix and
feeding the weighted sums through activations is called a {feature map}.
Importantly, the weights connecting the two layers must be the same, regardless
of where the kernel is located. Thus it is as though the kernel window is
scanning the input matrix for smaller features which are translationally
invariant. For example, in number recognition, the network may learn to
associate rounded edges with a zero. What these features are in reality relies
on the weights learned during training. A single convolution layer will usually
use several feature maps to generate the input for the next layer. 

\subsection*{Overfitting}
To improve a network's predicting power against unseen data, overfitting must be
avoided. Overfitting occurs during training when accuracy against the training
dataset continues to grow but accuracy against unseen data stops improving.
The network is not learning general features of the data anymore. This occurs
when the complexity of the net architecture has more computing potential than
required. The opposite problem is underfitting, using too small a network which
is incapable of learning data to high accuracy.

An obvious solution to overfitting is early stopping, cutting the training short
once accuracy against unseen data ceases to improve. However, we also wish to
delay overfitting such that this accuracy is as large as possible after
stopping.

In this paper we also make use of dropout to avoid overfitting. Dropout is a
technique where neurons in a given layer have a probability of being switched off
during one training round. This forces neurons to learn more general features
about the dataset and can decrease overfitting \citep{Dropout}.

\section{Brief Overview of Support Vector Machines}\label{SVM}
\begin{figure}
	\centering
	\includegraphics[width=0.46\textwidth]{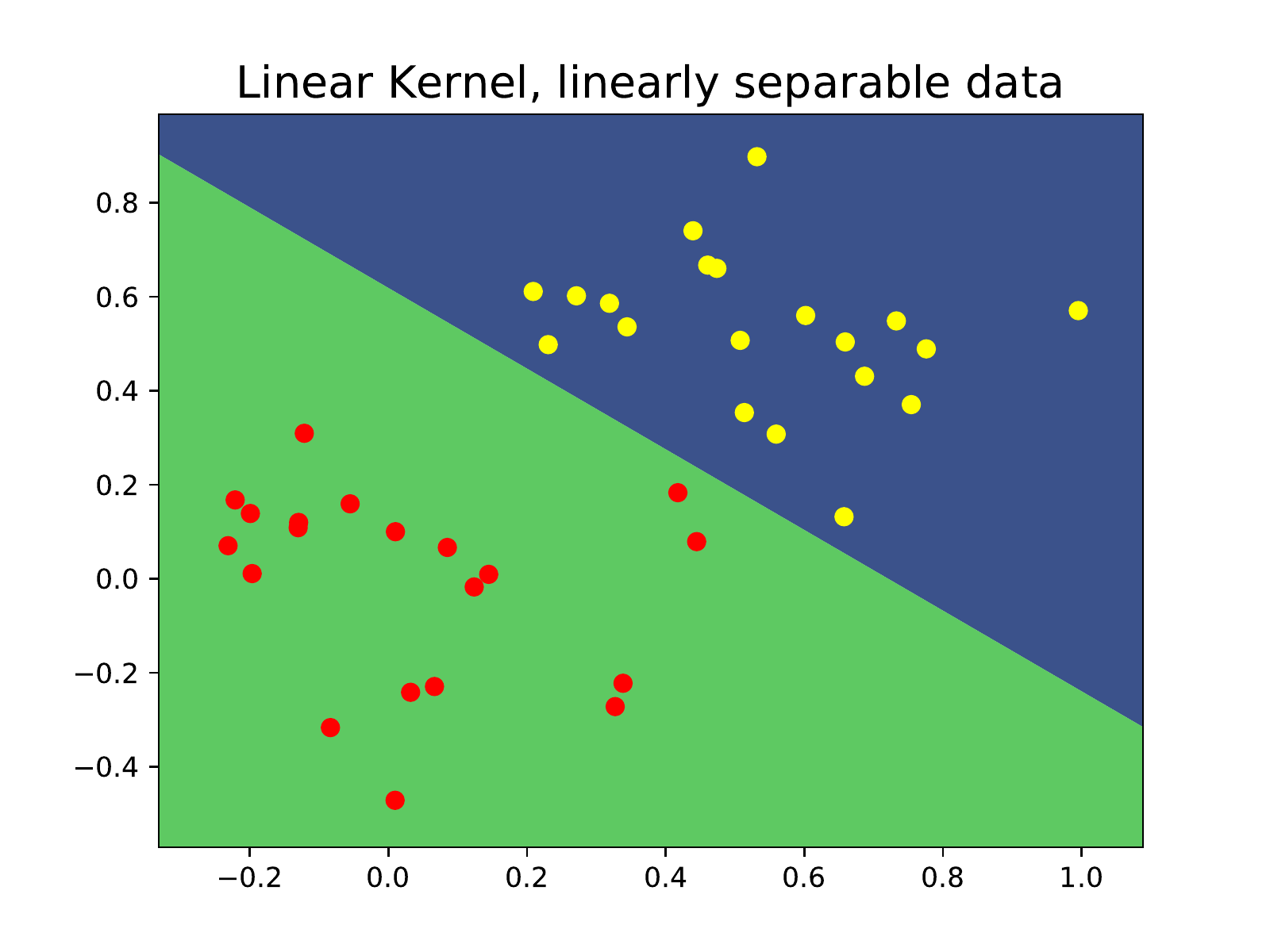}
	\includegraphics[width=0.46\textwidth]{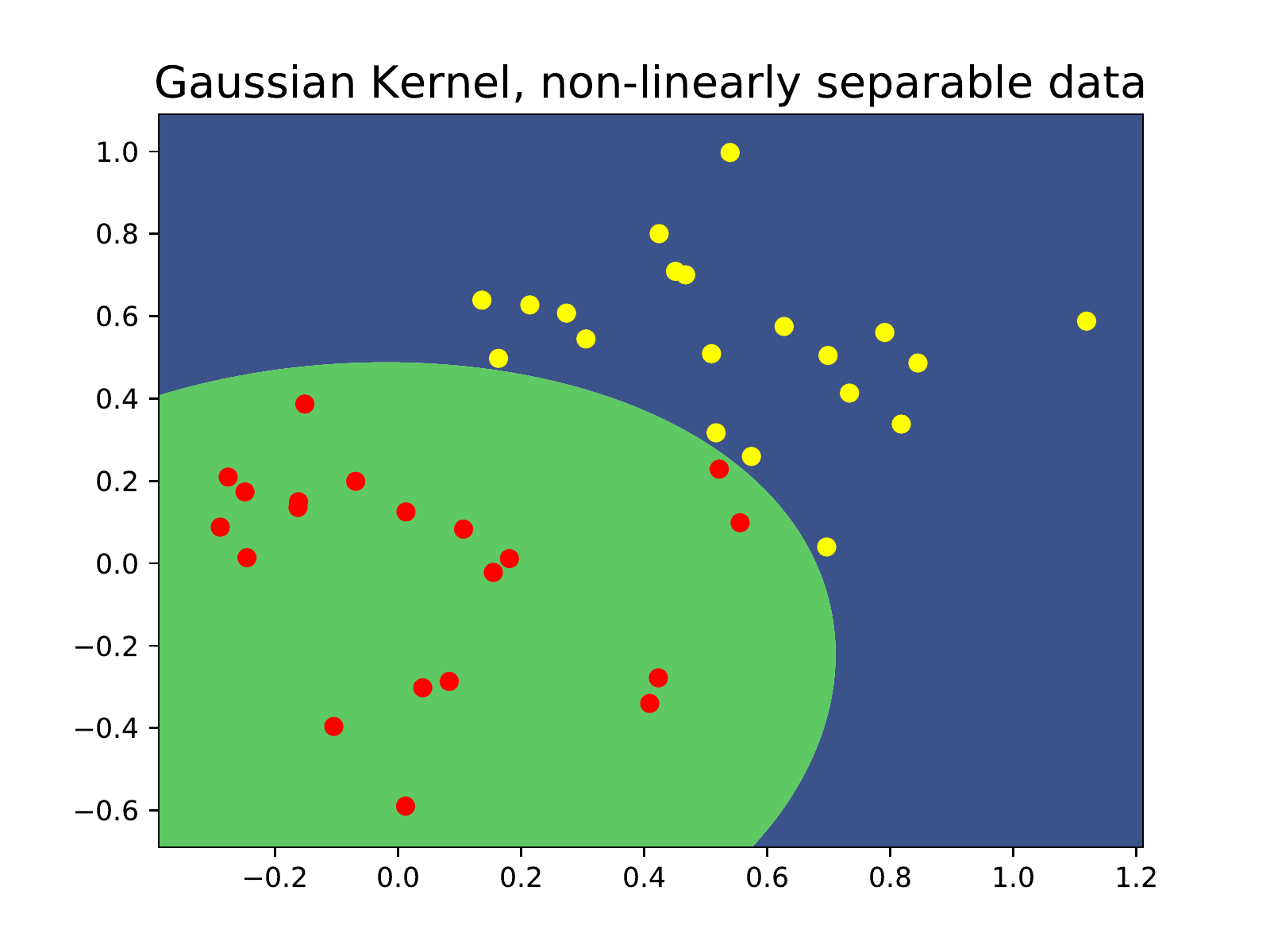}
	\caption{{\sf Example SVM separation boundary calculated using our Cvxopt implementation with a randomly generated data set.}}
	\label{fig:ex_svm}
\end{figure}

In contrast to neural networks, support vector machines (SVMs) take a more
geometric approach. They can act as both classifiers and regressors, but it is
more instructive to begin this discussion about classifiers. While a neural
network classifier essentially fits a large number of parameters (weights and
biases) to obtain a desired function $f(\mathbf{v_{in}})=0,1$, a SVM tries to
establish an optimal hyperplane separating clusters of points in the
{feature space} (the $n$-dimensional Euclidean space to which the
$n$-dimensional input vector belongs).  Points lying on one side of the plane are
identified with one class, and vice versa for the other class.  Thus a vanilla
SVM is only capable of acting as a binary classifier for linearly separable
data. This is somewhat restrictive, but the approach can be generalised to
non-linearly separable data via the so called \textit{kernel trick} and likewise can be
extended to deal with multiple classes \citep{svm} (see Figure \ref{fig:ex_svm}).

We wish to separate points $\mathbf{x_i}$ with a hyperplane based on a classification
of true/false, which we represent with the labelling $y_i=\pm1$. First define a hyperplane
\begin{equation}
	\left\{ x \in \mathbb{R}^n | f(\mathbf{x}) = \mathbf{w \cdot x} + b = 0
	\right\},
	\label{eq:hplane_def}
\end{equation}
where $\mathbf{w}$ is the normal vector to the hyperplane.
{Support vectors} are the points in the feature space lying closest to
the hyperplane on either side which we denote as $\mathbf{x^{\pm}_i}$. Define
the {margin} as the distance between these two vectors projected along
the normal $\mathbf{w}$, \textit{i.e.}, $\text{Margin} := \mathbf{w \cdot
(x^{+}_i-x^{-}_i})/|\mathbf{w}|$.
There is typically not a unique hyperplane we could choose to separate labelled
points in the feature space, but the most optimal hyperplane is one which
maximises the margin. This is because it is more desirable to have points
lie as far from the separating plane as possible, as points close to the
boundary could be easily misclassified. Note that condition defining a
hyperplane (\ref{eq:hplane_def}) is not unique as a rescaling $\alpha(\mathbf{w
\cdot x}+b)=0$ describes the same hyperplane. Thus we can rescale the normal
vector such that $f(\mathbf{x^{\pm}_i})=\pm 1$ and the margin reduces to
\begin{align}
	\text{Margin} = \frac{2}{|\mathbf{w}|}.
\end{align}
Moreover, with such a rescaling, the SVM acts as a classifier on an input
$\mathbf{x}$ through the function $\text{sgn}(f(\mathbf{x}))=\pm 1$.
Maximising the margin thus corresponds to minimising $|\mathbf{w}|$, with the
constraint that each point is correctly classified. This wholly defines the
problem which can be stated as
\begin{align}
	&\text{Min    } 
	\frac{1}{2} |\mathbf{w}|^2  
	\text{    subject to    }\  
	y_i(\mathbf{w} \cdot \mathbf{x_i}+b) \geq 1.
\end{align}
This is a quadratic programming problem with well known algorithms to 
solve it. Reformulating this problem with Lagrange multipliers:
\begin{align}
	L &= \frac{1}{2} |\mathbf{w}|^2 - \sum_i 
	\alpha_i (y_i(\mathbf{w} \cdot \mathbf{x_i} + b) - 1) ,\\
	\frac{\partial L}{\partial \mathbf{w}} &= \mathbf{w} -
	\sum_i \alpha_i y_i \mathbf{x_i} = 0 , \quad
	\frac{\partial L}{\partial b} =
	-\sum_i \alpha_i y_i = 0, \nonumber
\end{align}
leads to the dual problem:
\begin{align}
	&\text{\textbf{Dual:} Min} \quad
	\frac{1}{2} \sum_{i,j} 
	\alpha_i \alpha_j y_i y_j \mathbf{x_i} \cdot \mathbf{x_j}
	 - \sum_j \alpha_j , \quad\\
	 &
	\quad\text{subject to} \ 
	\alpha_j \geq 0 \text{,} \quad \sum_j \alpha_j y_j =0. \nonumber
\end{align}
With our classifying function now being $\text{sgn}(f(\mathbf{x})) = \text{sgn}
\left( \sum_i \left( \alpha_i y_i \mathbf{x_i \cdot x} \right) + b \right)$.
Again this is a quadratic programming problem. In this study we solve the dual
problem by using the Python package {Cvxopt}, which implements a
quadratic computing algorithm to solve such problems.

The dual approach is much more illuminating as it turns out the only $\alpha_i$
which are non-zero correspond to the support vectors \citep{svm} (hence the name
support vector machine). This makes SVMs rather efficient as unlike a neural
net which requires a vast amount of parameters to be tuned, a SVM is fully
specified by its support vectors and is ignorant of the rest of the data.
Moreover the quadratic programming optimisation implemented via Cvxopt ensures
the minimum found is a global one.

The dual approach also enables us to generalise to non-linearly separable data
rather trivially. In theory, this is achieved by mapping points in the feature
space into a higher dimensional feature space where the points are
linearly-separable, finding the optimal hyperplane and then mapping back into
the original feature space. However in the dual approach, only the dot product
between vectors in the feature space is used. Thus in practice we can avoid the
mapping procedure as we only need the effective dot product in the higher
dimensional space, known as a {kernel}. Thus by replacing $\mathbf{x_i
\cdot x}$ with $\text{Ker}(\mathbf{x_i,x})$ we can deal with non-linearly
separable data at almost no extra computational cost. This is known as the
\textit{kernel trick}. Common kernels include:
\begin{align}
	\text {Gaussian:    }  \text{Ker}(\mathbf{x_i,x}) &= 	\exp \left(\frac{-|\mathbf{x_i-x}|^2}{2 \sigma} \right) , \\
	\text {Polynomial:    }  \text{Ker}(\mathbf{x_i,x}) &= \left(1+\mathbf{x_i \cdot x}\right)^n.\nonumber
\end{align}

In our study of CICYs we exclusively use the Gaussian kernel as this leads to
the best results. SVMs can also act as a linear regressor by finding a function $f(\mathbf{x}) =
\mathbf{w \cdot x} + b$ to fit to the data. Analogous to the above discussion,
one can frame this as an optimisation problem by choosing the flattest line
which fits the data within an allowed residue $\epsilon$. Likewise one can make use of
Lagrange multipliers and the kernel trick to act as a non linear regressor too.

Note in the above discussion we have avoided the concept of {slack}. In
order to avoid overfitting to the training data, one can allow a few points in
the training data to be misclassified in order to not constrain the hypersurface
too much, allowing for better generalisation to unseen data. In practice this
becomes quantified by replacing the condition $\alpha_i \geq 0$ with $0 \leq
\alpha_i \leq C$, where $C$ is the {cost} variable \citep{svm}. 

\subsection*{SVM Regressors} \label{svm_reg}
The optimisation problem for a linear SVM regressor follows from finding the flattest
possible function $f(\mathbf{x}) =\mathbf{w \cdot x} + b$ which fits the data within a
residue $\epsilon$. As $|\nabla f |^2 = |\mathbf{w}|^2$, this flatness condition
reduces to the problem:
\begin{align}
	&\text{Min    } 
	\frac{1}{2} |\mathbf{w}|^2  
	\text{    subject to    } 
	-\epsilon \leq y_i - (\mathbf{w \cdot x_i} + b) \leq \epsilon
\end{align}
Introducing Lagrange multipliers
\begin{align}
	L &= \frac{1}{2} |\mathbf{w}|^2 - \sum_i 
	\alpha_i (y_i-(\mathbf{w} \cdot \mathbf{x_i} + b) + \epsilon) \nonumber \\
        &+\sum_i \alpha_i^\star (y_i-(\mathbf{w} \cdot \mathbf{x_i} + b) - \epsilon)\,, \\
	\frac{\partial L}{\partial \mathbf{w}} &= \mathbf{w} -
	\sum_i (\alpha_i - \alpha_i^*) y_i \mathbf{x_i} = 0 , \quad
	\frac{\partial L}{\partial b} =
	\sum_i (\alpha_i-\alpha_i^*) y_i = 0, \nonumber
\end{align}
leading to the dual problem:
\begin{align}
	\text{\textbf{Dual:} Min    } 
	&\frac{1}{2} \sum_{i,j} 
	(\alpha_i-\alpha_i^*) (\alpha_j-\alpha_j^*) y_i y_j \ \mathbf{x_i} \cdot
\mathbf{x_j} 
+\epsilon \sum_i \left(\alpha_i+\alpha_i^* \right) \nonumber \\
	&+ \sum_i  y_i(\alpha_i^*-\alpha_i) 
\end{align}
subject to the conditions 
\begin{equation}
\alpha_i,\alpha_i^* \geq 0 \text{,} \sum_i (\alpha_i-\alpha_i^*) =0.
\end{equation}

Thus, identical to the classifier case, this optimisation problem can be
implemented with Cvxopt. As the dual problem again only contains a dot product
between two entries in the feature space, we can use the kernel trick to
generalise this approach to fit non-linear functions.

\section{Hyperparameter Optimisation} \label{hyper}
\begin{figure}[h]
	\centering
	\includegraphics[width=0.94\textwidth]{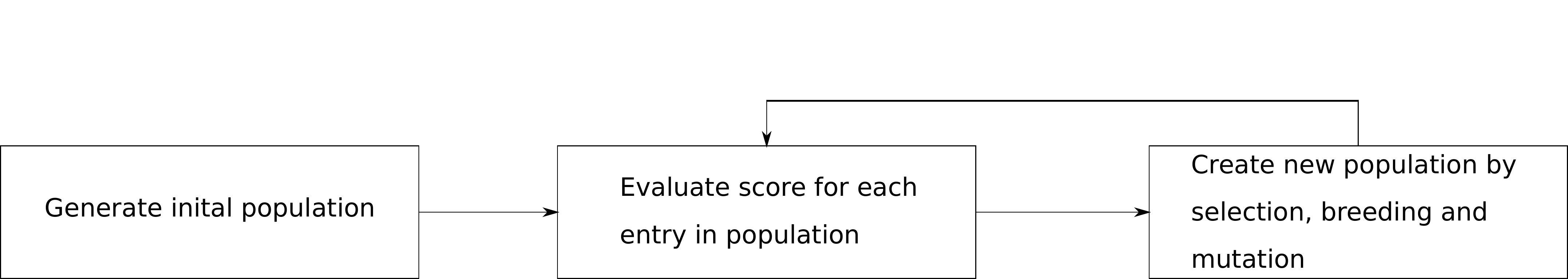}
	\vspace{5pt}
	\caption{{\sf Schematic of general genetic algorithm.}}
	\label{fig:gen_schem}
	\vspace{10pt}
\end{figure}
While both neural networks and SVMs are trained algorithmically as outlined in
Appendices \ref{NeuralNetworks} and \ref{SVM}, certain variables must be set by
hand prior to training. These are known as \textit{hyperparameters}. Examples
include net architechture (number of hidden layers and neurons in them) and
dropout rate for feedforward neural networks, kernel size and number of feature
maps for convolution layers and the cost variable, kernel type and kernel
parameters for SVMs.

Several methods exist to optimise these parameters. For the
case of a few hyperparameters, one could search by hand, varying parameters
explicitly and training repeatedly until an optimal accuracy is achieved. A
grid search could also be used, where each parameter is scanned through a range
of values. However, for a large number of hyperparameters permitting a large
number of values this quickly becomes an extremely time consuming task. Random
searches can often speed up this process, where parameter values are drawn from
a random distribution across a sensible range. In this study we make use of a
genetic algorithm, which effectively begins as a random search but then makes an
informed decision of how to \textit{mutate} parameters to increase accuracy.

In general, a genetic algorithm evolves a population of \textit{creatures}, each
creature having associated with it a list of parameters and a score. Each new
generation consists of the top scorers from the previous generation and
\textit{children} which are \textit{bred} from the last generation.
More specifically, to breed better models we create an initial
population of models, each being initialised with a random set of
hyperparameters. Each model is trained to its early stopping point and its
validation accuracy recorded. The models with the top $20$\% (arbitrary) validation
accuracy are kept, along with a few others by chance.

\textit{Breeding} then consists of pairing the surviving models into parents and
forming new models by choosing each hyperparameter randomly from one of the
parents.  Bred models also have a small chance to randomly \textit{mutate} its
parameters.
For a genetic algorithm to be successful, it is crucial to allow mutations and a
few low scoring nets into the next generation. This ensures there is enough
variance in the parameters available as to avoid a local minima.

\bibliographystyle{JHEP} 
\bibliography{bibfile}


\end{document}